\documentclass[11pt]{article}
\usepackage{a4wide}
\usepackage{amsmath,amssymb,amsfonts}
\usepackage{epsfig,verbatim,graphics}
\usepackage{subfigure,slashed}

\usepackage{tikz}
\usetikzlibrary{arrows,shapes}
\usetikzlibrary{trees}
\usetikzlibrary{matrix,arrows} 				
\usetikzlibrary{positioning}				
\usetikzlibrary{calc,through}				
\usetikzlibrary{decorations.pathreplacing}  
\usepackage{pgffor}							

\usetikzlibrary{decorations.pathmorphing}	
\usetikzlibrary{decorations.markings}
\tikzset{
    vector/.style={decorate, decoration={snake}, draw},
	provector/.style={decorate, decoration={snake,amplitude=2.5pt}, draw},
	antivector/.style={decorate, decoration={snake,amplitude=-2.5pt}, draw},
    fermion/.style={draw=black, postaction={decorate},
        decoration={markings,mark=at position .55 with {\arrow[draw=black]{>}}}},
    fermionbar/.style={draw=black, postaction={decorate},
        decoration={markings,mark=at position .55 with {\arrow[draw=black]{<}}}},
    fermionnoarrow/.style={draw=black},
    gluon/.style={decorate, draw=black,
        decoration={coil,amplitude=4pt, segment length=5pt}},
    scalar/.style={dashed,draw=black, postaction={decorate},
        decoration={markings,mark=at position .55 with {\arrow[draw=black]{>}}}},
    scalarbar/.style={dashed,draw=black, postaction={decorate},
        decoration={markings,mark=at position .55 with {\arrow[draw=black]{<}}}},
    scalarnoarrow/.style={dashed,draw=black},
    electron/.style={draw=black, postaction={decorate},
        decoration={markings,mark=at position .55 with {\arrow[draw=black]{>}}}},
	bigvector/.style={decorate, decoration={snake,amplitude=4pt}, draw},
}

\tikzstyle{block} = [draw, rectangle, 
    minimum height=3em, minimum width=6em]

\newcommand{\mathsym}[1]{{}}

\newcommand{\eref}[1]{(\ref{#1})}

\renewcommand\({\left(}
\renewcommand\){\right)}
\renewcommand\[{\left[}
\renewcommand\]{\right]}

\newcommand{\dd}{{\rm d}}
\declareslashed{}{{^-}}{.1}{0.1}{\rm d}
\newcommand{\dbar}{ {\slashed{\rm d}}}
\newcommand{\e}{{\rm e}}

\newcommand\vp{\varphi}
\newcommand\eps{\epsilon}

\newcommand\mn{{\mu\nu}}

\def\ba{\begin{eqnarray}}
\def\ea{\end{eqnarray}}
\def\be{\begin{equation}}
\def\ee{\end{equation}}

\def\mcL{\mathcal{L}}
\def\D{\mathcal{D}}
\def\H{\mathcal{H}}

\def\nn{\nonumber}
\def\({\left(}
\def\){\right)}

\def\eref#1{(\ref{#1})}

\newcommand{\roughly}[1]{\mathrel{\raise.3ex\hbox{$#1$\kern-0.85em
\lower1ex\hbox{$\sim$}}}}

\begin{document}

\begin{titlepage}
\begin{center}

\today
\hfill\phantom{.}

\vskip 1.5cm

{\LARGE \bf Effective action for the Abelian Higgs model in FLRW}

\vskip 1cm

{\bf Damien P. George\footnote{dpgeorge@nikhef.nl}, Sander Mooij\footnote{smooij@nikhef.nl} and Marieke  Postma\footnote{mpostma@nikhef.nl}}

\vskip 25pt

{\em
\hskip -.1truecm  Nikhef, \\Science Park 105, \\1098 XG Amsterdam,\\ The
Netherlands \vskip 5pt }

\end{center}

\vskip 0.5cm

\begin{center}
{\bf ABSTRACT}\\[3ex]
\end{center}
We compute the divergent contributions to the one-loop  action of the U(1) Abelian Higgs model.  The calculation allows for a
Friedmann-Lema\^itre-Robertson-Walker space-time and
a time-dependent expectation value for the scalar field.  Treating the
time-dependent masses as two-point interactions, we use the in-in
formalism to compute the first, second and third order graphs that
contribute quadratic and logarithmic divergences to the effective
scalar action.  Working in $R_\xi$ gauge we show that the result is
gauge invariant upon using the equations of motion.

\end{titlepage}

\newpage \setcounter{page}{1} \tableofcontents

\newpage


\section{Introduction}

There is only one scalar field in the standard model, but it plays a crucial
role.  Scalars are more abundant in most extensions of the standard
model, such as the multiple Higgs fields in grand unified theories,
the superpartners in supersymmetric models, and the moduli fields in
extra-dimensional set-ups.  To study the physics of the early universe
and to test physics beyond the standard model using cosmological data,
it is important to have a precise understanding of the scalar field's
dynamics. This requires to go beyond a classical treatment and include
the dominant quantum effects.

The one-loop effective action for a scalar field in an expanding
universe has been known for a long time
\cite{birrell,candelas,ringwald1,ringwald2}.  It describes the backreaction of the quantum fluctuations of the scalar field on the
(time-dependent) background field, which can be calculated
systematically in a loop expansion.  If the scalar is coupled to other
scalars or to fermions via e.g. a Yukawa interaction, additional
scalar and fermion loops contribute
\cite{candelas,greene,heitmannfermion}.  In this paper we extend these
results by including a coupling to a gauge field.  That is, we calculate
the effective action for a Higgs-like field, which is charged under a
gauge symmetry, in a Friedmann-Lema\^itre-Robertson-Walker (FLRW) universe. For
simplicity we focus on an Abelian symmetry, but the results are easily
generalized to more general gauge groups.  This is the generalization
of the Coleman-Weinberg potential \cite{CW} to time-dependent
background fields in a curved space-time.

Our results are of direct importance for inflation models in which the
SM Higgs or a GUT Higgs is the inflaton (or a waterfall) field
(e.g. \cite{Bezrukov:2007ep,rachel2}). The Coleman-Weinberg potential
gives the dominant quantum correction during inflation.  The
time-dependent corrections may become important at the end of
inflation, and during the subsequent period of reheating.  Another
application is the description of flat directions of the MSSM and its
extensions \cite{flatland}, which are lifted by the one-loop quantum
corrections.  This may affect inflation models or Affleck-Dine
baryogenesis models \cite{AD1,AD2} using flat directions.

The effective action for an Abelian gauge theory in de Sitter
space-time has been calculated by \cite{shore,allen,ishikawa} using
the Landau gauge.  More recently the calculation was done in the
$R_\xi$ gauge, showing gauge invariance of the effective action
\cite{garbrecht}.  To obtain this result an adiabatic approximation
was made which fails in the $\xi \to 0$ limit.  We extend these
results to a generic FLRW space-time and allow for the possibility of
time-dependence of the background field, which in a cosmological
set-up can be displaced from its potential minimum.  The calculation
is done in the $R_\xi$ gauge using a perturbative approach. We
calculate the quadratic and logarithmic divergent terms in the
ultraviolet (UV) limit, which come from a finite number of diagrams
(and which do not depend on the perturbation being small throughout).
The resulting effective action is gauge invariant only on-shell, that
is after using the classical equations of motion, in agreement with the
Nielsen identities \cite{fukuda,nielsen}.

Our results agree with the expressions in the literature in the
appropriate limit. In the limit of a static background field and a
constant Hubble parameter our results agree with \cite{garbrecht}.  In
the Minkowski limit we retrieve the effective action calculated in our
previous work \cite{MP}, and also the effective equations of
motion found earlier in
\cite{heitmann1,heitmann2,heitmann3,boyanovsky}.  Finally, taking both
a static background field and a static background we get the familiar
Coleman-Weinberg potential \cite{CW}.

To properly take into account the real-time evolution of the system,
and to assure the results are manifestly real, we use the in-in
formalism (also known as closed-time-path (CTP) or Schwinger-Keldysh
formalism) to calculate the effective action
\cite{ctpschwinger,ctpkeldysh,ctpjordan,Bakshi1,Bakshi2,ctpcalzetta,ctpweinberg}. We
first derive the one-loop corrected equation of motion for the Higgs
field, using the tadpole method \cite{tadpole}.  All one-loop Feynman-diagrams with
one external Higgs leg contribute. For technical reasons it is easier
to work in conformal time, as the resulting action has a form
similar to the Minkowski action, and all the machinery developed for
this \cite{MP,heitmannmt} can be used. We split all two-point interactions into
time-independent and time-dependent parts, treating the former as masses and the latter as
interaction terms; in the loop-expansion the result will not depend on
this split \cite{CW}.  The equations of motion can be formally integrated to
obtain the effective action up to field-independent terms \cite{Heitmannphdt}.  Finally,
we can rewrite the results in coordinate time.

The effective action is independent of the specific initial conditions
chosen.  We will argue that this is always the case, for arbitrary
initial conditions, provided the initial vacuum is chosen to be that of
the free theory. The different vacua can be related via a Bogoliubov
transformation \cite{heitmanninitial}.

As already mentioned, we only calculate the UV divergent terms, as
these will generically give the dominant contribution.  Using a
renormalization prescription, these terms (together with the
wavefunction renormalization of the gauge field) suffice to derive the
renormalization group equations (RGE) and find the RG improved action.
We neglect the backreaction on space-time, and treat the FLRW scale
factor as classical background field.  
Finally, we note that to apply the results to models of Higgs
inflation, a non-minimal coupling to gravity has to be considered.
All this is left however for future work.

In the next subsection we give a self-contained summary of the
results.  Following this we go through the calculation, starting
in Sec.~\ref{s:action_and_formalism} describing the model, the
in-in formalism, and giving the vertices and propagators needed to
compute Feynman diagrams.  In Sec.~\ref{s:eom} we calculate the
relevant graphs at first, second and third order which contribute
to the one-loop equation of motion.  These graphs are used in
Sec.~\ref{s:effective} to compute the effective action for the
charged scalar.  Here we also present results when additional
scalar and fermions run in the loop.  Our choice of initial
conditions, and their generalization, is discussed in
Sec.~\ref{s:bogol}.  We conclude in Sec.~\ref{s:concl}, and
provide some further details of the calculation in a pair of
appendices.

\subsection{Summary of the results}

Here we shall outline the model and give the main result of
our calculation.  It is self-contained so that one need not get
caught up in the details of the derivation to make use of the
final answer.

The ansatz for the space-time metric is 
\be
\dd s^2 = \dd t^2 - a^2(t)\dd \vec x^2, 
\ee
with $a(t)$ the time-dependent scale factor of the FLRW metric.  The
action is that of an Abelian Higgs model, with a Higgs field charged
under a U(1) gauge symmetry
\be
S_{\rm tot} = \int \dd^4 x\sqrt{-g} \left[
-\frac14 g^{\mu \alpha} g^{\nu \beta} 
F_{\mu \nu} F_{\alpha \beta} 
+g^{\mu \nu} D_\mu \Phi (D_\nu \Phi)^\dagger -
V(\Phi) \right].
\ee
We expand the Higgs into a time-dependent background field (the
zero-mode) $\phi$ plus quantum fluctuations $h$ and $\theta$
\be
\Phi(x^\mu) = \frac{1}{\sqrt{2}}\big( \phi(t) +h(t,\vec x)+ i \theta(t,\vec x)\big).
\label{expand}
\ee
We gauge fix using $R_\xi$ gauge (see details in next section).  The
effective potential is corrected by loops which are calculated using
the in-in formalism.  We sum over all the relevant loops, and in the
end go on-shell, which ensures the gauge invariance of the final result.
We find, up to background field-independent terms, the UV divergent
contributions at one loop to be
\be
\Gamma^{\rm 1-loop} \!=\! \frac{-1}{16\pi^2}
\!\int\! \dd^3 x \dd t \sqrt{-g}\Bigg[
(\tilde{V}_{hh} + \tilde{V}_{\theta \theta} + 3 m_A^2)\Lambda^2
- \(
\tilde{V}^2_{hh}+\tilde{V}^2_{\theta\theta}
+3 m_A^4 -6 \tilde{V}_{\theta\theta}  m_A^2
\) \frac{\ln (\Lambda/\bar{m})^2}{4}  \Bigg],
\ee
where the ``shifted scalar mass'' is 
\be
\tilde{V}_{\alpha\alpha}\equiv
V_{\alpha\alpha}-\dot H-2H^2,
\label{mass_shift1}
\ee
which is time-dependent.  A subscript on $V$ denotes a derivative with
respect to that field.
Further, $\Lambda$ is the cutoff used in regulating the divergent
momentum integrals (it cuts off 3-momentum $|\vec k| < \Lambda$),
and the arbitrary mass $\bar{m}$ is put in to ensure the argument
of the logarithm is dimensionless (recall that we are only interested
in divergent contributions to the effective action).
$H=\dot{a}/a$ is the Hubble constant, with a dot denoting a derivative
with respect to time $t$.  The (time-dependent) mass of the
gauge field is $m_A^2=g^2\phi^2$. 

Our result agrees with those found in the literature.  For the
Minkowski case ($H=\dot H=0$, and thus $\tilde V _{\alpha \alpha} =
V_{\alpha\alpha}$) it matches our previous result \cite{MP}.  In
the de Sitter limit $\dot H =0$, and for a time-independent Higgs field
($V_{\theta\theta} =0$ by Goldstone's theorem), it agrees with
Garbrecht \cite{garbrecht}.  Finally, taking both the Minkowski limit
and a static background field, we retrieve the classic
Coleman-Weinberg potential \cite{CW}.

If the Higgs field couples to additional scalars $\chi_\alpha$ and/or
fermion fields $\psi_\beta$,\footnote{We assume here a Yukawa
  interaction $m_\psi \propto \phi$. The expression for a more general
  mass term is given by \eref{gamma_fermion}. We work in  a basis
  where both the mass and kinetic terms are diagonal; this can be
  easily generalized.} we get an additional
contribution
\begin{align}
\delta \Gamma^{1-{\rm loop}} &=
-\frac{1}{16\pi^2}
\int \dd^3 x \dd t \sqrt{-g}\Bigg[ \sum_{\chi_\alpha} \(
\tilde{V}_{\alpha\alpha} \Lambda^2
-
\tilde{V}_{\alpha\alpha}^2
 \frac{\ln (\Lambda/\bar{m})^2}{4}  \)
\nn \\
&\qquad - \sum_{\psi_\beta} \(
m^2_{\beta} \Lambda^2
-
\(m^4_{\beta} - \tilde{V}_{\theta\theta} m_\beta^2\)
 \frac{\ln (\Lambda/\bar{m})^2}{4}  \)
\Bigg],
\end{align}
where the shifted scalar mass is given by \eref{mass_shift1}.
Here the sum is over all bosonic and fermion real degrees of
freedom, where a Weyl (Dirac) fermion counts as 2 (4) degrees
of freedom.

\section{Action and formalism}
\label{s:action_and_formalism}

The starting point for our calculation is the action of a U(1)
Abelian-Higgs model in an FLRW background.  The background space-time
is fixed, in the sense that the backreaction of the charged scalar
is assumed negligible.  We work in $R_\xi$ gauge, and therefore include
in the action a gauge fixing and a Faddeev-Popov term.  We work with
a conformal metric, where most expressions take a form reminiscent
of the Minkowski calculation.  Since the background space-time, as
well as the scalar vacuum expectation value (its classical value),
are taken to be time-dependent, the ``masses'' (really two-point
interactions) of the particles are also time-dependent.  We deal
with this by splitting these two-point interactions into a
time-independent part, which we call the mass and which determines
the propagator, and a time-dependent part, which is treated as a
proper two-point interaction in Feynman diagrams.
The action, propagators and interaction vertices are defined in
the following subsections, along with the in-in formalism for computing
expectation values.  We use this machinery in the next section to
calculate the one-loop equation of motion.

\subsection{Notation}

We use a metric with signature $(+,-,-,-)$, indexed by lower Greek
letters $\mu,\nu\ldots$, and lower Latin letters for just the 3-space.
The Greek letters $\alpha,\beta,\ldots$, and $I,J$, index the set of
quantum fields.
For propagators, covariant derivatives, mode function normalization
etc.\ we use the same conventions as Peskin and Schroeder~\cite{peskinschroeder}.
Masses $m$ and frequencies $\omega$ are split into time-independent and
time-dependent parts, with the notation $m^2(t) = \bar m^2 + \delta m^2(t)$,
with $\delta m^2(0)=0$.  A hat above a mass scale denotes the
corresponding quantity rescaled by the scale factor: $\hat m = a
m$.  Derivatives with respect to conformal time $\tau$ are denoted by
a prime, and derivatives with respect to coordinate or physical time
$t$ by a dot. In momentum integrals $\dbar k = \dd k/(2\pi)$.

\subsection{The action in an FLRW background}
\label{s:action}

The FLRW metric in physical and conformal coordinates is, respectively,
\be
\dd s^2 = \dd t^2 - a^2(t)\dd \vec x^2
= a^2(\tau) \( \dd \tau^2 - \dd \vec x^2\).
\label{metric}
\ee
The non-zero connections are
\be
\Gamma^i_{i0} = \Gamma^i_{0i} = \Gamma^0_{00} = \Gamma^0_{ii} = \H .
\label{Gamma}
\ee
Here we defined $\H = a'/a$, analogous to the usual definition in
coordinate time $H = \dot a/a$.  We can decompose the charged
scalar field into a real and imaginary part,
\be
\Phi(x^\mu) = \frac{1}{\sqrt{2}}\big( \phi(\tau) +h(\tau,\vec x)+ i \theta(\tau,\vec x)\big) ,
\label{phi_decompose}
\ee
with $\phi(\tau)$ the time-dependent classical background field.  The
action is a sum of the kinetic and potential terms, the gauge fixing term and
the Faddeev-Popov term:
\be
S_{\rm tot} = \int \dd^4 x\sqrt{-g} (\mcL+ \mcL_{\rm GF} + \mcL_{\rm FP}),
\label{action}
\ee
with
\begin{align}
\mcL &= -\frac14 g^{\mu \alpha} g^{\nu \beta} 
F_{\mu \nu} F_{\alpha \beta} 
+g^{\mu \nu} D_\mu \Phi (D_\nu \Phi)^\dagger -
V(\Phi) ,\\
\mcL_{\rm GF} &= -\frac1{2\xi} G^2, \qquad G = g^{\mu\nu}\nabla_\mu A_\nu -\xi
g(\phi+h) \theta, \\
\mcL_{\rm FP} &= {\bar \eta} g \frac{\delta G}{\delta \alpha} \eta.
\end{align}
Note that $\nabla_\mu g^{\mu \nu} =0$ (because of metric
compatibility), and thus $g^{\mu\nu}\nabla_\mu A_\nu =\nabla_\mu
g^{\mu\nu} A_\nu$ and there is no ambiguity.  $\delta G/\delta \alpha$
is the operator obtained by computing the variation of $G$ under a U(1)
gauge transformation with infinitesimal parameter $\alpha$.
A bar over $\eta$ denotes the conjugate, not to be confused
with a time-independent quantity.

Our aim is to compute the quantum corrected equation of motion for
the background field $\phi$.  It can be derived by demanding that all
tadpole diagrams with one external $h$-leg vanish.
To calculate these diagrams we need to derive both the
propagators, which follow from the free and time-independent part
of the action, and the interaction vertices, which include
time-dependent two-point interactions, as well as three-point
interactions.\footnote{There are also four-point interactions but
these do not contribute to the one-loop tadpoles.}
The key to our approach, following \cite{heitmannmt}, is the choice to separate out from the
two-point terms the constant, diagonal pieces.  This allows
us to easily solve for the scalar and gauge propagators, since the
masses are then really masses: they are field-diagonal and time
independent.  The left-over pieces
are treated as two-point interaction vertices, and put into graphs
following the usual Feynman rules.

To make explicit all factors of the scale factor we now write
$g^{\mu \nu} = a^{-2} \eta^{\mu \nu}$ with $\eta^{\mu\nu}$
the Minkowski metric, which is the metric in the comoving frame with
coordinates $(\tau, \vec x)$.  In the expressions
below, all indices are raised and lowered using the Minkowski metric.
We denote all mass scales in comoving coordinates with a hat. In
particular we define:
\be
\hat \phi_\alpha = a \phi_\alpha, \qquad \hat V = a^4 V(\hat \phi),
\label{hat}
\ee
with $\phi_\alpha=\{\phi,h,\theta,\eta\}$ the scalars in the theory.  The
hatted fields are canonically normalized in the comoving frame.  Since
the gauge field kinetic terms are conformally invariant, there is no
rescaling of the gauge field.  These comoving fields feel a potential
$\hat V$.  All the comoving quantities map directly to the equivalent
set-up in Minkowski, and we can use the usual Minkowski machinery to
calculate Feynman diagrams.

The action \eref{action} is expanded in quantum fluctuations around
the background.  Here we state the results at each order; for details
see Appendix~\ref{A:action}.  The one-point vertex is \eref{S1}
\be
S^{(1)}=
\int \dd^4 x \( -\hat \lambda_h \hat h \),
\ee
where
\be
\hat \lambda_{h} =  \( \partial_\tau^2 -(\H'+\H^2) \)\hat \phi +
\hat V_{\hat \phi}
= a^3\[ \ddot \phi +3 H \dot \phi + V_\phi\] .
\label{l1}
\ee

The quadratic action is \eref{S2}
\begin{align}
S^{(2)}&=
\int \dd^4 x \Bigg\{
-\frac12 A_\mu \[ -(\partial^2+\hat m^2_{(\mu)}) \eta^{\mu\nu} +
\(1-\frac1{\xi}\) \partial^\mu \partial^\nu \]A_\nu
-A_0 (\hat m^2)^{i0} A_i
\nn \\
& \hspace{2cm}
-\hat m^2_{A\theta} A_0 \hat \theta 
-\frac12  \sum_{\vp_i = \{\hat h,\hat \theta\}}
\hat \vp_i  ( \partial^2 + \hat m^2_{\vp_i} )\hat \vp_i
-\hat{\bar{\eta}}  ( \partial^2 + \hat m^2_{\eta} ) \hat \eta
\Bigg\}.
\label{S2_in_text}
\end{align}
The explicit form of the two-point interactions, with $\hat m^2 = a^2 m^2$, are:
\begin{align}
&
\hat m^2_{(\mu)} = g^2 \hat \phi^2 +\frac2{\xi}\( \H' -2 \H^2
\) \delta_{\mu0}  = a^2\[ g^2 \phi^2 + \frac{2}{\xi}
\(\dot H - H^2 \) \delta_{\mu0}  \],
\nn\\
&
\hat m^2_h = \hat V_{hh} - (\H'+\H^2)= a^2\[ V_{hh}
-(\dot H +2 H^2)\],
 \nn\\
&
\hat m^2_\theta = \hat V_{\theta \theta} + \xi g^2 \hat \phi^2
-(\H'+\H^2)
=a^2\[  V_{\theta \theta} + \xi g^2\phi^2-(\dot H +2 H^2)\],
\nn \\
&
\hat m^2_\eta = \xi g^2 \hat \phi^2 -(\H'+\H^2)
=a^2\[\xi g^2 \phi^2 -(\dot H +2 H^2)\],
\label{mass}
\end{align}
where we used $\H^2 =a^2 H^2$ and $\H' = a^2(\dot H + H^2)$.
The off-diagonal two-point terms are:
\be
\hat m^2_{A\theta} = 2 g (\partial_\tau -\H)\hat \phi = a^2\[2g\dot \phi\], 
\qquad (\hat m^2)^{0i} =   (\hat m^2)^{i0}
= \frac{2}{\xi} \H \partial^i.
\ee
The mixing between the spatial $A_i$ and temporal $A_0$ gauge field
contains a first derivative, which survives in all gauges except for
unitary gauge $\xi \to \infty$.  However, we will not work in unitary
gauge as this is known to give false results if not done carefully (see for example \cite{GK}),
and we are thus forced to deal with this extra complication.

As stated above, we choose to split the two-point interactions into a
time-independent and time-dependent part:
\be
\hat m^2(\tau) = \hat{\bar m}^2 +\delta \hat m^2(\tau).
\label{mass_split}
\ee
The first term contributes to the free Lagrangian from which the propagator
is constructed, whereas the time-dependent term is treated as a
two-point interaction.  The loop expansion is independent of the split
of the two-point terms into a free and interacting part \cite{CW}.  The split
is defined by requiring the interaction to vanish at the initial time,
which we choose without loss of generality to be at $t_0 =0$:
\be
\delta m^2(0) =0.
\label{BC0}
\ee
For the diagonal two-point interactions, \eref{BC0} serves only to define
$\bar m^2$ unambiguously, and does not constrain anything physically
meaningful, like, for example, the initial conditions of the background
scalar and scale factor.  However, due to the presence of off-diagonal
two-point interactions, and the fact that we want to retain a Minkowski-like
propagator structure to simplify the calculation, we will need to
make some choice.  In particular, we choose initial conditions such
that the initial off-diagonal two-point interactions and Lorentz violating
contribution to the gauge boson mass (the term proportional to
$\delta_{\mu0}$) vanish at the initial time: $m^2_\text{off-diag}(0)=0$.
This implies the initial conditions
\be
\delta \phi(0) = \delta \phi'(0) = \H(0) = \H'(0)=0,
\label{BC1}
\ee
where we wrote $\phi(t) = \bar \phi + \delta \phi(t)$.  As we will
argue in Sec.~\ref{s:bogol}, the results we obtain do not depend on the
specific initial conditions chosen, provided we define the initial vacuum
as that of the free theory.  It is therefore no real limitation that the boundary
conditions above are not the most physically motivated ones in actual
cosmological settings.

The two-point interactions for the scalars, $\hat m_h^2$, $\hat m_\theta^2$ and
$\hat m_\eta^2$, are split as per \eref{mass_split}.  The diagonal gauge boson
two-point interaction is split as a constant degenerate piece, a time-dependent
degenerate piece, and an extra piece for $A^0$ denoted by $m_0^2$:
\be 
{\hat m}^2_{(\mu)} =
    \hat{\bar{m}}_A^2
    + \delta{\hat m}_A^2
    + \delta{\hat m}_0^2 \, \delta_{0\mu},
\label{massA}
\ee
with 
\be
\delta{\hat m}_0^2 = \frac2\xi(\H' -2 \H^2).
\label{mass0}
\ee
The off-diagonal two-point interactions do not have a time-independent
part, so $\delta\hat m_{A\theta}^2=\hat m_{A\theta}^2$ and
$(\delta\hat m^2)^{0i}=(\hat m^2)^{0i}$.  For these terms, we shall
use the notation with and without the $\delta$ interchangeably.

The three-point interaction vertices are \eref{S3}
\be
S^{(3)} =
\int \dd^4 x \[
    -\frac12 \hat \lambda_{h\alpha\alpha} \hat h (\hat \psi_\alpha)^2
    -\hat \lambda_{h\eta\eta} \hat h \hat {\bar\eta} \hat\eta
    -\hat h \hat \lambda_{h A\theta} A_0 \hat\theta
\] ,
\ee
where $\psi_\alpha$ runs over the quantum fields $\{h,\theta,A\}$, and
\begin{align}
\hat\lambda_{hhh} &=
    \partial_{\hat\phi} \hat m_{h}^2
    = \hat V_{\phi hh}, \nn \\
\hat\lambda_{h\theta\theta} &=
    \partial_{\hat\phi} \hat m_{\theta}^2
    = \hat V_{\phi\theta\theta} + 2 \xi g^2 \hat\phi, \nn \\
\hat\lambda_{hAA} &=
    \partial_{\hat\phi} \hat m_A^2
    = -2g^2 \hat\phi, \nn \\
\hat\lambda_{h\eta\eta} &=
    \partial_{\hat\phi} \hat m_\eta^2
    = 2 \xi g^2 \hat\phi, \nn \\
\hat\lambda_{hA\theta} &= 2g(-\partial_\tau-\H).
\label{lambda}
\end{align}
Note that $\hat\lambda_{h A\theta}$ contains a derivative of
conformal time, which, by using partial integration of \eref{S3}, we
let act on the factor $A_0 \hat\theta$ instead of $\hat h$.  This
allows us to factor out a common $\hat h$ in $S^{(3)}$ and thereby
compute the tadpole diagrams.\footnote{Equation~\eref{S3} also contains
a term $-2g A^i\hat\theta\partial_i \hat h$.  Since the final expression
of each tadpole graph is independent of the spatial coordinates, this
three-point interaction does not contribute to the overall result.  We
have checked this by explicit computation.}
In \eref{lambda} we express the diagonal couplings
$\hat\lambda_{h\alpha\alpha}$ as derivatives, with respect to $\hat\phi$,
of the corresponding two-point interaction defined by \eref{mass}.
In computing these derivatives, we have assumed a fixed FLRW
background so that $\H$ is independent of $\hat\phi$ and
$\partial_{\hat\phi} \H'=\partial_{\hat\phi} \H^2=0$.

\subsection{In-in formalism and propagators}
\label{s:propagators}

Since we are interested in expectation values of the background field,
and their evolution with time, rather than scattering amplitudes, we
will use the in-in formalism (also known as closed-time-path or
Schwinger-Keldysh formalism) \cite{ctpschwinger, ctpkeldysh,ctpjordan, Bakshi1,Bakshi2,ctpcalzetta, ctpweinberg}.
In contrast with the usual in-out formalism, this approach gives
results that are manifestly real.

Expectation values are computed using an action $S = S[\phi_i^+]
-S[\phi_i^-]$, with boundary condition $\phi_i^+(t) = \phi_i^-(t)$.
That is, we double the fields, and take the action for the plus-fields as
for the minus-fields, given by the equations in the previous subsection.
All fields, propagators and vertices are labeled by $\pm$ superscripts.
In a Feynman diagram the
propagator $D^{\pm\pm}(x-x')$ connects between a $\lambda^\pm(x)$ and
a $\lambda^{\pm}(x')$ vertex.
Since the action of the minus-fields is defined with an overall minus
sign we have
\be
\[m^2_{\alpha\beta}\]^- = -\[m^2_{\alpha\beta}\]^+, \qquad
\lambda_{h\alpha\beta}^- = -\lambda_{h\alpha\beta}^+.
\label{minus}
\ee

We construct the propagators from the free, time-independent part of the
quadratic action \eref{S2_in_text}.  The corresponding quadratic Lagrangian
can be written in the form
\be
\mcL^{\rm free}[\phi_\alpha^+]=-(1/2)\sum_{\alpha,\beta} \phi^+_\alpha(x^\mu) 
\bar K^{\alpha\beta} (x^\mu) \phi_\beta^+(x^\mu),
\ee
with the sum over all fields $\phi_\alpha = \{h,\theta,\eta,A^\mu\}$.
For example, the scalar fields have
$\bar K^{\alpha\beta} = (\partial^2 + \bar m^2)\delta^{\alpha\beta}$.
The propagators are then defined as the solutions of
\be
\( \begin{array}{cc}
\bar K^{\alpha\beta}(x^\mu) & 0 \\
0 & -\bar K^{\alpha\beta}(x^\mu)
\end{array} \)
\( \begin{array}{cc}
D_{\beta\gamma}^{++}(x^\mu - y^\mu) & \;D_{\beta\gamma}^{+-}(x^\mu - y^\mu) \\
D_{\beta\gamma}^{-+}(x^\mu - y^\mu) & \;D_{\beta\gamma}^{--}(x^\mu - y^\mu)
\end{array} \)
= -i \delta^\alpha_{\gamma} \delta(x^\mu - y^\mu) \mathbf{I}_{2}.
\label{defG}
\ee
This defines $D^{++}$ as the usual Feynman propagator, $D^{--}$ as the
anti-Feynman propagator, and $D^{-+}$ and $D^{+-}$ as Wightman
functions.
Our initial conditions (\ref{BC0},\ref{BC1}) are such that the
FLRW propagators in conformal coordinates are analogous to the usual Minkowski
expressions.

It turns out convenient to rewrite all the propagators in terms of
Wightman functions. We introduce the shorthand for the Wightman
function
\be
D_{I,ab} \equiv D_I^{-+}(x_a-x_b), \quad
D_{\mu\nu, ab} \equiv D_{\mu\nu}^{-+}(x_a-x_b),
\label{Dshort}
\ee
for the propagator of a type-$I$ scalar, and the gauge boson
propagator, respectively.  Using this the propagators are
(suppressing the $I$ or the Lorentz indices)
\begin{align}
D^{++}(x_a -x_b) &= D_{ab} \Theta_{ab}+ D_{ba} \Theta_{ba}, \nn \\
D^{--}(x_a-x_b) &= D_{ab} \Theta_{ba}+ D_{ba} \Theta_{ab}, \nn \\
D^{-+}(x_a-x_b) &= D^{+-}(x_b-x_a) = D_{ab},
\label{Dall}
\end{align}
with $\Theta_{ab} = \Theta(\tau_a-\tau_b)$ the usual step function.
We Fourier transform the propagator with respect to comoving three-momentum
\be 
D(x_a-x_b) = \int \dbar^3 k D(\tau_a-\tau_b,\vec k)
    \e^{i \vec k \cdot (\vec x_a - \vec x_b)},
\label{D3k}
\ee
with $\dbar k = \dd k/(2\pi)$.
The time dependence $(\tau_a-\tau_b)$ in the Fourier propagator
will from now on be suppressed.
For a scalar field the solution for the Fourier transformed
Wightman function is
\be
D_I(\vec k) = \frac{1}{2\hat{\bar \omega}_I(k)}
    \e^{-i \hat{\bar \omega}_I(k) (\tau_a-\tau_b)},
\label{Dh}
\ee
where $\hat{\bar \omega}_I^2(k)=\hat{\bar m}_I^2+{\hat k}^2$
and $\hat{\bar m}_I^2$
is the (conformal) mass-squared of the appropriate scalar.
The Fourier transformed Wightman function for the gauge boson
propagator is
\be
D_{\mu\nu}(\vec k) = -\(\eta_{\mu\nu} -\frac{{\hat k}_\mu {\hat k}_\nu}{\hat{\bar m}_A^2}\) D_A(\vec k)
-\xi \frac{{\hat k}_\mu {\hat k}_\nu}{\hat{\bar m}_\xi^2} D _\xi(\vec k) ,
\label{DA}
\ee
with $D_A$ and $D_\xi$ scalar propagators with mass-squared
$\bar m_A^2$ and $\bar m_\xi^2 = \xi \bar m_A^2$ respectively.  In the
$\xi =1$ gauge the gauge boson propagator is diagonal, $D_{\mu\nu} =
-\eta_{\mu\nu}D_A$.

The mixed two-point interaction $(\delta{\hat m}^2)^{0i}$ contains
a spatial derivative; it acts on the gauge boson propagator by
\begin{align}
(\delta{\hat m}^2)^{0i}(\tau_a) D_{i\mu,ab}(\vec k) 
&= - i {\hat k}^i \frac{2}{\xi} \H(\tau_a)  D_{i\mu,ab}(\vec k), 
\nn \\
(\delta{\hat m}^2)^{0i}(\tau_a)  D_{i\mu,ab}(\vec k) 
&= -(\delta{\hat m}^2)^{0i}(\tau_b)  D_{i\mu,ab}(\vec k) ,
\label{actionmix}
\end{align}
where we note that $(\delta{\hat m}^2)^{i0 }$ and $D_{\mu\nu,ab}$ are
diagonal in the Lorentz indices.  The boundary conditions (\ref{BC0},
\ref{BC1}) imply that $(\hat {\bar m}^2)^{0i} =\hat{\bar m}^2_{A\theta}
=\hat{\bar m}^2_{0}=0$; the off-diagonal two-point interactions and the
Lorentz violating $A_0$ mass contribution only enter the interaction Lagrangian.

\section{One-loop equation of motion}
\label{s:eom}

The equation of motion for the background Higgs field $\phi(t)$ follows
from the vanishing of the tadpole.  In terms of diagrams, these are
all one-particle irreducible tadpole graphs with one external $h^+$ leg.
In this section we compute these diagrams, and thus the quantum
corrected equation of motion, at the one-loop level.  We are here
concerned only with the UV divergent contributions to the graphs, and
thus need to consider only the diagrams with up to three vertices.
Throughout this work we use a cutoff regularization scheme for the momentum integrals. For the goal of this work, this seems the most intuitive approach, since the momentum integrals are over three-momentum $\vec{k}$, and we cut off $|\vec{k}|<\Lambda$.  Other regularization methods, such as for instance dimensional regularization, would give equivalent answers.
The calculation is done in the conformal frame, in terms of hatted
fields and mass scales, conformal time and momenta.  For notational
convenience, in this section we drop the hat on all quantities; it shall
be reinstated at the end when we give the results.

The calculation is analogous to the one for Minkowski \cite{MP}, but
with two-point interactions \eref{mass} that now depend on the FLRW scale
factor. This is straightforward to incorporate for the diagrams with a
scalar running in the loop.  There are however some new technical
difficulties that come in with the gauge boson loops:
\begin{enumerate}
\item The mass of the temporal gauge boson $m_{0}^2$ gets FLRW
  corrections but the mass of the spatial components $m_{i}^2$ does
  not.  This is possible because Lorentz symmetry is broken by the
  time-dependent background.  Consequently the diagrams with $A_0$ and
  $A_i$ contribute differently.
\item The off-diagonal gauge boson two-point interaction $(\delta m^2)^{0i}$
  is non-zero.  This results in new diagrams with both two and three
  two-point insertions. 
\item The formalism is set up in such a way that the two-point interactions vanish
  at the initial time \eref{BC0}. This avoids divergences that depend on the
  initial conditions. We will argue in Sec.~\ref{s:bogol} that this is
  always an allowed choice, for arbitrary initial conditions, provided the
  initial vacuum is chosen accordingly.
\end{enumerate}

The one-loop equation of motion can be extracted from the series of
tadpole diagrams with one external $h^+$ leg, see Fig.~\ref{F:tadpole}.
This gives
\begin{align}
0 &= A_{\rm cl} +A_1 +A_2 +A_3 +{\rm finite} 
\nn \\
&= \lambda^+_h(x) + \sum  \lambda^+_{h \alpha \beta}(x)\bigg [S_{\alpha\beta}
D^{++}_{\alpha \beta} (0)
-i S_{\alpha\beta\gamma\delta} \int\! \dd^4 x'
D^{+\pm}_{\alpha \gamma}(x-x')  \[\delta m_{\gamma\delta}^2 (x') \]^{\pm} D^{\pm +}_{\delta
  \beta}(x'-x)
\nn \\
&\quad-S_{\alpha\beta\gamma\delta\rho\sigma}\!\! \int\!\! \dd^4 x'\!\! \int\!\! \dd^4 x'' 
D^{+\pm}_{\alpha \gamma}(x-x')
\[\delta m_{\gamma\delta}^2(x')\]^\pm D^{\pm \pm}_{\delta \rho} (x'-x'')
\[\delta m_{\rho\sigma}^2(x'')\]^\pm D^{\pm +}_{\sigma\beta}(x''-x)
\nn \\
&\quad
+{\rm finite} \bigg].
\end{align}
where $A_i$ labels the $i^\text{th}$ order contribution to all tadpole
diagrams with $i$ vertices.  We labeled the classical contribution
$A_{\rm cl}$, which comes from a tree-level diagram.  Indices
$\{\alpha,\beta,\ldots\}$ are compound indices denoting field-type
as well as possible Lorentz indices for the gauge field.  The sum
is over these compound indices (all fields and their Lorentz indices)
as well as all possibilities for $\pm$. The $S_{\alpha\beta\ldots}$ are
appropriate symmetry factors, derived in Appendix \ref{A:tadpole}.
$\delta m_{\alpha\beta}^2$ and $\lambda_{h\alpha\beta}$ are the two-
and three-point interaction vertices, respectively, as defined in
Sec.~\ref{s:action}.

\begin{figure}[t]
\begin{center}
\begin{tikzpicture}
[line width=1.5 pt, scale=1.8]
\node at (-1,0) {$\sum A_i =$};
\begin{scope}[shift={(0.5,0)}]
\draw (0.0,0)--(1,0);
\node at (-0.2,0.1) {$ h^+$};
\draw[fill=black] (1,0) circle (.05cm);
\node at (.9,0.25) {$ \lambda_h$};
\end{scope}
\node at (3.5,0.) {$+$};
\begin{scope}[shift={(4.5,0)}]
\draw (0.0,0)--(1,0);
\node at (-0.2,0.1) {$ h^+$};
\draw (1.5,0) circle (0.5cm);
\draw[fill=black] (1,0) circle (.05cm);
\node at (.7,0.2) {$\lambda_{h\alpha\beta}$};
\node at (2.3,0.) {$D_{\alpha\beta}$};
\end{scope}
\begin{scope}[shift={(0,-2)}]
\node at (-0.7,0.0) {$+$};
\draw (0.0,0)--(1,0);
\node at (-0.2,0.1) {$ h^+$};
\draw (1.5,0) circle (0.5cm);
\draw[fill=black] (1,0) circle (.05cm);
\draw[fill=black] (2,0) circle (.05cm);
\node at (.7,0.2) {$ \lambda_{h\alpha\beta}$};
\node at (2.35,0.) {$\delta  m^2_{\rho\sigma}$};
\node at (3.5,0.) {$+$};
\node at (1.5,.7) {$D_{\alpha \rho}$};
\node at (1.5,-.7) {$D_{\sigma\beta}$};
\begin{scope}[shift={(4.5,0)}]
\draw (0.0,0)--(1,0);
\node at (-0.2,0.1) {$ h^+$};
\draw (1.5,0) circle (0.5cm);
\draw[fill=black] (1,0) circle (.05cm);
\draw[fill=black] (1.7,0.45) circle (.05cm);
\draw[fill=black] (1.7,-0.45) circle (.05cm);
\node at (.7,0.2) {$ \lambda^{\alpha\beta}$};
\node at (2.3,0.) {$D_{\sigma \kappa}$};
\node at (1.1,.7) {$D_{\alpha\rho}$};
\node at (1.1,-.6) {$D_{\beta \tau}$};
\node at (1.9,.65) {$\delta m^2_{\rho \sigma}$};
\node at (1.9,-.65) {$\delta m^2_{\kappa\tau}$};
\end{scope}
\end{scope}
\end{tikzpicture}
\caption{
\label{F:tadpole}
  Tree-level tadpole giving the classical equation of motion and the
  first, second and third order diagrams respectively.  The summation is over
  all fields, for the gauge bosons also over Lorentz indices, and over
  $\pm$ at the two-point vertices.}
\end{center}
\end{figure}

The tree-level tadpole diagram contributes \eref{l1}, and we recover
the classical equations of motion (remember we dropped the hat for
conformal coordinates and scales):
\be
0 = \lambda^{+}_h = \phi''  - (\H'+\H^2) \phi +V_{\phi}.
\label{eom_clas}
\ee
Given the equation of motion, we want to find the corresponding
action.  We do this by simply writing down an action which, upon
using the Euler-Lagrange equations, yields the equation of motion
\eref{eom_clas}.  This action is then transformed from the comoving
to the physical frame, thereby obtaining the effective action.
This will be done in Sec.~\ref{s:effective}.  The idea is to apply
this procedure to the quantum corrected equations of motion in order to
determine the quantum corrected effective action.

We divide the calculation based on the order of the contributing graphs,
which is the number of vertices in the loop of the tadpole.
As discussed above, we must work to third order.  Independent
of this, we can distinguish three classes of diagrams depending on how
they contribute to the answer.  First there is the
contribution that is fully analogous to the Minkowski calculation
$A^{\rm Mink} = A^{\rm Mink}_1+A^{\rm Mink}_2$, the only difference is
that the mass term of the scalars now depends on the scale factor.
Second is $A^{\rm mass}= A^{\rm mass}_2$, which arises from the extra
Feynman diagrams due to the FLRW mass correction of the temporal gauge
field $\delta m^2_0$; see \eref{massA}.
And finally $ A^{\rm mix}=A^{\rm mix}_2 + A^{\rm mix}_3$ gives the
diagrams with one and two off-diagonal vertices
$(\delta m^2)^{0i}$ connecting the temporal and spatial gauge fields,
also absent in Minkowski.

\subsection{First order contribution $A_1$ }

The calculation of the first order diagrams proceeds analogously to the
equivalent calculation in Minkowski, which can be found in \cite{MP}.
At first order, four diagrams contribute, with $\psi_\alpha=\{ h, \theta,
\eta,A^\mu\}$ running in the loop. The result only
depends on the time-independent part of the two-point interaction, as there is no
vertex insertion.  For each diagram the result has the same structure,
given by \eref{a1}
\be
(A_1^\text{Mink})_\alpha = \frac12 \partial_\phi m^2_\alpha D_\alpha^{++}(0) .
\ee
Just as in Minkowski the gauge loop can be expressed in terms of
scalar propagators \eref{DA} via
\be 
-\eta^\mn  D_{\mu \nu}^{++}(0)= 
3  D_A^{++}(0) + \xi  D_\xi^{++}(0).
\ee
The sum of all first-order diagrams is
\begin{align}
 A_1^{\rm Mink} 
&= \frac12\sum_\alpha \partial_{ \phi}  m^2_{\alpha}  D^{++}_\alpha(0)
=\frac12\sum_\alpha S_\alpha \partial_{ \phi}  m^2_{\alpha}
 \frac{1}{4\pi^2} \int_0^\Lambda k^2\dd k  \[ \frac{1}{k} -\frac12 \frac{ {\bar
     m}_\alpha^2}{k^3} +...\]
\nn\\
&=\frac1{16\pi^2}\sum_\alpha S_\alpha \partial_{ \phi}  m^2_{\alpha} 
\[ \Lambda^2 - \frac12 {\bar m}_\alpha^2 \ln  (\Lambda/\bar{m})^2 +{\rm finite} \].
\label{A1mink} 
\end{align}
In the momentum integrals here and below, the variable $k$ is the comoving momentum, $\Lambda$ is a comoving cutoff, and we have $k< \Lambda$. (Recall that graphs in this section in the comoving frame, and all quantities are actually hatted quantites.  The cutoff regularisation we apply here is equivalent to a physical cutoff on physical momentum.)
The sum is over $\alpha = \{h,\theta,\eta,A,\xi\}$, and $S_\alpha = \{1,1,-2,3,1\}$
counting the real degrees of freedom (with a minus sign for the
anti-commuting ghost). Further $m_A^2 =g^2 \phi^2$ and $m_\xi^2 =\xi
m_A^2$.  Note that the factor $\partial_\alpha m_\alpha^2$ is
time-dependent, and evaluated at $\tau$; hence $A_1^\text{Mink}$ is a
function of $\tau$.
The finite terms that we have neglected remain finite as $\Lambda\to\infty$.

\subsection{Second order contribution $A_2$}

At second order the loop diagrams with one two-point insertion contribute.  We
split them into three parts.  The first part, $A_2^{\rm Mink}$, contains all
scalar loops, and the gauge boson loop where only the diagonal part of 
\eref{massA} is inserted.  In addition there is a mixed $\theta
A^0$-loop \eref{a2theta}.  This part is analogous to the equivalent
Minkowski calculation.  The second part $A_2^{\rm mass}$ contains the
gauge boson loop with a $\delta m^2_0$, and the third part $A_2^{\rm mix}$,
with a $(\delta m^2)^{0i}$.  These last two diagrams are both absent in Minkowski.

\subsubsection{$A_2^{\rm Mink}$}

The scalar Higgs loop with one two-point insertion gives \eref{a2diagonal}
\begin{align}
A^{\rm Mink}_{2,h} &= -\frac{i}{2}\partial_{\phi} 
m^2_h(\tau_a) \int \dd^4 x_b \delta
m_h^2(\tau_b) \bigg[
D_h^{++} (x_a\!-\!x_b) D_h^{++} (x_b\!-\!x_a) 
\nn \\
&\qquad\qquad
- D_h^{+-} (x_a\!-\!x_b) D_h^{-+} (x_b\!-\!x_a) \bigg]
\nn \\
&=
-\frac{i}{2}\partial_{\phi} 
m^2_h(\tau_a) \int \dd^4 x_b \delta
m_h^2(\tau_b) \Theta_{ab}\bigg[
D_{h,ab}^2 - D_{h,ba}^2\bigg] ,
\end{align}
where we expressed the result in terms of Wightman functions, using
the notation introduced in Sec.~\ref{s:propagators}.  Fourier transforming
as in \eref{D3k} and performing the $\dd^3 x$ integral gives a
$\delta^3 (\vec k +\vec p)$.  And thus:
\begin{align}
A^{\rm Mink}_{2,h} &=-\frac{i}{2}\partial_{\phi} 
m^2_h(\tau_a) \int_0^{\tau_a} \dd \tau_b \delta
m_h^2(\tau_b) \int \dbar^3 k \bigg[
D_{h,ab} (\vec k) D_{h,ab} (\vec p) 
- D_{h,ba} (\vec k) D_{h,ba} (\vec p) \bigg]_{ \vec k
  = - \vec p}
\nn \\
&= -\partial_{\phi} 
m^2_h(\tau_a) \int_0^{\tau_a} \dd \tau_b \delta
m_h^2(\tau_b) \int \frac{\dbar^3 k}{(2\bar
  \omega_h)^2}
\sin\[ 2\bar \omega_h (\tau_a-\tau_b)\] ,
\end{align}
where we used the explicit form of the propagator \eref{Dh}.
Now use integration by parts to extract the UV divergent piece (we
give here the general formula, which can also be applied to the gauge
boson loop discussed below)
\be
\int_0^{\tau_a}\! \dd \tau_b  f(\tau_b) \int\!
\frac{\dbar^3 k }{(2\bar \omega_I) (2\bar \omega_J)} 
\sin[(\bar \omega_I+\bar \omega_J) (\tau_a-\tau_b)]
= f(\tau_a)\int\! \dbar^3 k\frac{1} {(2\bar \omega_I) (2\bar \omega_J)
  (\bar \omega_I+\bar \omega_J)} +... 
\label {A2partial}
\ee
where we assumed $f(0)=0$ and the ellipses denote higher order terms
${\mathcal O}(\vec k^{-4})$. Putting it all back together we find for
the scalar loop
\begin{align}
A^{\rm Mink}_{2,h} &= -\partial_{\phi} 
m^2_h(\tau) \delta
m_h^2(\tau) \int \frac{\dbar^3 k}{(2\bar \omega_h)^3 }
=-\partial_{\phi} 
m^2_h(\tau) \delta
m_h^2(\tau) \int \frac{\dbar^3 k}{8 \vec k^3}
\nn \\
&=-\partial_{\phi} 
m^2_h(\tau) \delta
m_h^2(\tau)  \frac{1}{32\pi^2}\ln (\Lambda/\bar{m}) ^2 + \text{finite},
\end{align}
where we rewrote the time variable $\tau_a = \tau$.
To get to the second expression we expanded in large momentum.

The Goldstone boson and ghost loops give a similar contribution,
although for the ghost with an overall factor $(-2)$ to take into
account that these are two real anti-commuting degrees of freedom.  The
calculation of the gauge boson loop follows the same steps, except
that now care has to be taken of the Lorentz structure.  We find
\begin{align}
A^{\rm Mink}_{2,A} &= -\frac{i}{2}\partial_{\phi} 
m^2_A \int \dd^4 x_b \delta
m_A^2(\tau_b) \eta^{\mu\nu} \eta^{\rho \sigma }\bigg[
D_{\mu\rho}^{++} (x_a-x_b) D_{\sigma\nu}^{++} (x_b-x_a) 
\nn \\ & \qquad
- D_{\mu\rho}^{+-} (x_a-x_b) D_{\sigma\nu}^{-+} (x_b-x_a) \bigg]
\nn \\
&= -\partial_{\phi} 
m^2_A \int_0^{\tau_a} \dd \tau_b \delta
m_A^2(\tau_b) \int \dbar^3 k \frac{C_{IJ}}{(2\bar
  \omega_I)(2\bar\omega_J)}
\sin\[ (\bar \omega_I + \bar \omega_J) (\tau_a-\tau_b)\]
\nn \\
&=-\partial_{\phi} 
m^2_A \delta
m_A^2(\tau)  \frac{(3+ \xi^2)}{32\pi^2}\ln (\Lambda/\bar{m})^2 +{\rm finite} .
\end{align}
Again, we rewrote $\tau_a=\tau$.
The relevant propagator combination, which defines $C_{IJ}$, is
\begin{align}
&\eta^{\mu\nu} \eta^{\rho\sigma} D_{\mu \rho}(\vec k) 
D_{\sigma  \nu}(\vec p) \big|_{\vec k = -\vec p} 
=C_{IJ}(k) D_I D_J
\\
  &\qquad =
\(3+\frac{4\vec k^2 \bar \omega_A^2}{\bar m_A^4}\) D_A(\vec k)^2
+\xi^2  \(1+\frac{4\vec k^2 \bar \omega_\xi^2}{\bar m_\xi^4}\)
D_\xi(\vec k)^2
-2\xi \frac{\vec k^2 (\bar \omega_A +\bar \omega_\xi)^2}{\bar m_A^2 \bar m_\xi^2} D_A(\vec k)
D_\xi(\vec k), \nn
\end{align}
with $I,J = A,\xi$, and $D_{I,ab} = (2\bar \omega_I)^{-1} \e^{-i\bar \omega_I(\tau_a-\tau_b)}$.

Finally, the mixed $\theta A^0$-loop gives \eref{a2theta}
\begin{align}
A^{\rm Mink}_{2,A\theta} &= -i  
\lambda_{h\theta A} (\tau_a)\int\dd^4 x_b \;
\delta m_{A\theta}^2(\tau_b) 
\[D_{00,ab}^{++} D_{\theta,ba}^{++}-D_{00,ab}^{+-} D_{\theta,ba}^{-+}\]
\nn \\
&= 
-2\lambda_{h\theta A} (\tau_a)\int_0^{\tau_a} \dd \tau_b \;
\delta m^2_{A\theta}(\tau_b)
\int \dbar^3 k
\sum \frac{C_I}{(2\bar \omega_I)(2\bar \omega_\theta)}
\sin[(\bar \omega_I+\bar \omega_\theta)(\tau_a-\tau_b)] 
\nn \\
&=
2\lambda_{h\theta A} (\tau) 
\delta m^2_{A\theta}(\tau)\frac{(3+\xi)}{128 \pi^2} \ln (\Lambda/\bar{m})^2 + \text{finite} ,
\end{align}
with $\lambda_{h\theta A}(\tau) =2g(-\partial_{\tau} -\H(\tau))$ the
appropriate three-point vertex.\footnote{Note that $\int \dd \tau
  \lambda_{h\theta A} (\tau) \delta m^2_{A\theta}(\tau) = \int \dd
  \tau\partial_\phi m^2_{A\theta} (\tau) \delta m^2_{A\theta}(\tau) $
  after integration by parts, which we will use in the expression for the
  effective action in Sec.~\ref{s:effective} to rewrite $(A^{\rm
    Mink}_2)_{A\theta}$ in the same form as the other contributions.}
In the second line we identified
\be
D_{00} = \sum C_I D_I = -\(1-\frac{\bar \omega_A^2}{\bar m_A^2}\) D_A
-\xi \frac{\bar \omega_\xi^2}{\bar m_\xi^2}D_\xi .
\ee
In the final step we performed integration by parts \eref{A2partial},
and took the large $\vec k$ limit.

Adding everything together gives
\be
 A^{\rm Mink}_2 =  -\frac{1}{32\pi^2} \sum_\alpha S_\alpha \partial_{ \phi} 
m^2_\alpha   \delta  m^2_\alpha
\ln  \Lambda^2
+
\frac{(3+\xi)}{64\pi^2} \lambda_{hA\theta}
 \delta  m^2_{A\theta}\ln (\Lambda/\bar{m})^2 +{\rm finite}.
\label{A2mink}
\ee
%

\subsubsection{$A_2^{\rm mass}$ and $A_2^{\rm mix}$}

The last two diagrams to contribute at second order are those with an
$m_0^2$ and an $(m^2)^{0i}$ insertion, the Lorentz violating mass and
off-diagonal gauge boson interaction respectively.  Neither diagram is present
in Minkowski. The loop with $m_0^2$ gives \eref{a2diagonal}
\begin{align}
 A_2^{\rm mass} &=
-\frac{i}{2}\partial_\phi m^2_A(\tau_a)
\int \dd^4 x_b \; \delta m_0^2(\tau_b)
\eta^{\mu\nu}
\[ D_{\mu0,ab}^{++} D_{0\nu,ba}^{++} - D_{\mu0,ab}^{+-} D_{0\nu,ba}^{-+} \]
\nn \\
&= -\partial_{\phi} 
m^2_A (\tau_a) \int_0^{\tau_a} \dd \tau_b \;
\delta m_{0}^2(\tau_b) 
\int \dbar^3 k
\sum \frac{C_{IJ}}{(2\bar \omega_I) (2\bar \omega_J)}
\sin\[(\bar \omega_I+\bar \omega_J)(\tau-\tau_b)\]
 \nn \\ 
&=
-\partial_{\phi} 
m^2_A(\tau) \delta m_{0}^2(\tau) \frac{(3+\xi^2)}{4\times 32\pi^2} \ln (\Lambda/\bar{m})^2 +
{\rm finite} .
\label{A2mass}
\end{align}
Here we defined the relevant propagator combination by
\begin{align}
&\eta^{00} \eta^{\mu\nu} D_{\mu 0}(\vec k) D_{0 \nu}(\vec p)\big|_{\vec
  k =- \vec p} 
=\sum  C_{IJ} D_I(\vec k)D_J(\vec k) 
\label{D_mass}\\
&= \frac{\vec k^2(2 \vec k^2+ \bar m_A^2)}{\bar m_A^4} D_A(\vec k) ^2
+\xi^2 \frac{(\bar \omega_\xi^2 + \vec k^2)\bar \omega_\xi^2}{\bar m_\xi^4}
D_\xi(\vec k)^2 
- 2\xi \frac{\vec k^2 \bar \omega_\xi(\bar \omega_\xi +\bar
  \omega_A)}{\bar m_A^2 \bar m_\xi^2} D_A(\vec k) D_\xi(\vec k) ,
\nn
\end{align}
with as before $I,J = A,\xi$, and $D_I(-\vec k) = D_I(\vec k)= 1/(2\bar
\omega_i) \e^{-i\bar \omega_I (\tau - \tau_b)}$.

The off-diagonal interaction $(\delta m^2)^{0i}$ contains a spatial derivative,
and brings down a factor of the momentum. The diagram is given by
\eref{a2mix}
\begin{align}
 A_2^{\rm mix} &=i \partial_{ \phi} 
m^2_A(\tau_a) \int \dd^4 x_b (\delta m^2)^{0i} (\tau_b)
\eta^{\mu\nu}
\bigg[ D^{++}_{\mu 0 ,ab}  D_{i \nu,ba}^{++} 
-D^{+-}_{\mu 0,ab} D_{i\nu,ba}^{-+} \bigg]
\nn \\
&=-\frac2{\xi} \partial_{ \phi} 
m^2_A(\tau_a) \int_0^{\tau_a}\! \dd \tau_b \H(\tau_b) \!\int\! \dbar^3 k \, p^i \eta^{\mu\nu}
\bigg[ D_{ab,\mu 0}(\vec k) D_{ab, i\nu}(\vec p) 
+D_{ba,\mu 0}(\vec k) D_{ba, i\nu}(\vec p) \bigg]_{\vec k =- \vec p}
\nn\\
&=-\frac{2}{\xi} 
\partial_{\phi} 
m^2_A(\tau_a) \int_0^{\tau_a} \dd \tau_b \H(\tau_b) \!\int\! \dbar^3 k \frac{C_{IJ}}{(2\bar
  \omega_I)(2\bar \omega_J)} 2\cos\[(\bar \omega_I + \bar \omega_J)(\tau_a-\tau_b)\]
\nn\\
&=-
\frac{2}{\xi}
\partial_{\phi} 
m^2_A(\tau_a) \int \dbar^3 k \frac{2 \H'(\tau_a)}{(2\bar
  \omega_I)(2\bar \omega_J)(\bar \omega_I+\bar \omega_J)^2} C_{IJ}
+{\rm finite}
\nn\\
&=
\partial_{\phi} 
m^2_A(\tau) \frac{3\H'(\tau)(1-\xi)^2}{64\pi^2 \xi} 
\ln (\Lambda/\bar{m})^2 +{\rm finite}.
\label{A2mix}
\end{align}
As we now have a cosine instead of a sine in the expression on
the third line above, we integrate by parts twice to isolate the leading
term in the UV limit.  This is why the result is proportional
to $\H'$.  The relevant propagator contribution is defined by
\begin{align}
&p^i \eta^{\mu\nu} D_{\mu0}(\vec k) D_{i\nu} (\vec p) 
\big]_{\vec k = -\vec p}  \equiv \sum C_{IJ} D_I D_J
\nn \\
&\qquad= 
-\frac{\vec k^2 \bar \omega_A (2\vec k^2+\bar m_A^2)}{\bar m_A^4}
D_A^2-\frac{\vec k^2 \bar \omega_\xi (2\vec k^2+\bar m_\xi^2)}{\bar m_A^4}
D_\xi^2
+\frac{\vec k^2(\bar \omega_A+\bar \omega_\xi)(\vec k^2+ \bar \omega_A \bar \omega_\xi)}{\bar m_A^4} D_A D_\xi .
\nn 
\end{align}

\subsection{Third order contribution $A_3$}

The third order diagrams with two two-point insertions are UV finite, which
can be easily checked by power counting.  The only exception to this
is the diagram with two off-diagonal $(m^2)^{0i}$ insertions, because
each insertion contains a spatial derivative, and thus brings down a
power of momentum. We thus consider the third order diagram with two
mixed-interaction insertions \eref{a3a}
\be
A^{\rm mix}_{3} =\frac12 \partial_{ \phi} 
m^2_A(\tau_a) \int \dd^4 x_b \dd^4 x_c 
(\delta m^2)^{0i} (\tau_b) (\delta m^2)^{0j}(x_c)
\sum_{\rho,a} \eta^{\mu\nu} D_{\mu\rho,ab}
D_{\sigma \kappa,bc} D_{\tau\nu,ca} ,
\label{M_1}
\ee
where the sum $\sum_{\rho,a}$ is over the four possibilities for
the Lorentz indices
\be
(\rho,\sigma,\kappa,\tau) =  (i,0,j,0), (0,i,0,j) ,(0,i,j,0),(i,0,0,j)
\label{indices}
\ee
and also over the four possibilities for plus and minus fields.
The $x_a$ vertex is a $+$-vertex.  There are then 4 possibilities
for the vertices:
\be
(a,b,c) = (+++),\; (+-+),\; (++-),\; (+--) .
\label{abc}
\ee
The propagator between a $(\pm)$-vertex and a $(\pm)$-vertex is
$D^{\pm\pm}$, and we write them out in terms of Fourier transformed
Wightman functions using the notation of section \eref{s:propagators}.
Taking the action of the spatial derivatives in $(m^2)^{0i}$ will
then bring down powers of momentum, as per \eref{actionmix}.
Finally, the $\vec x_b$ and $\vec x_c$ integrals give delta functions
encoding momentum conservation.  At the end of the day we find
\begin{align}
A_3^{\rm mix} &=  \frac2{\xi^2} \partial_{ \phi} 
m^2_A(\tau_a) \int_0^{\tau_a} \dd \tau_b \int_0^{\tau_a} \dd \tau_c \H (\tau_b)\H(\tau_c) 
\int \dbar^3 k \sum_\rho k^i k^j s_\rho
\\
&\times \Bigg[  \( D_{ba}(\vec k) D_{bc}(\vec p) D_{ac}(\vec q)
+ {\rm c.c} \)_{\vec k = \vec q =-\vec p}
- 2 \Theta_{bc}  \( D_{ab}(\vec k) D_{bc}(\vec p) D_{ac}(\vec q)
+ {\rm c.c} \)_{\vec k = \vec p =-\vec q} \bigg],
\nn
\end{align}
where the sum $\sum_\rho$ is now only over the Lorentz indices
\eref{indices}, which we suppressed in the above formula. The sign
$s_\rho = (1,1,-1,-1)$ for the four possibilities \eref{indices}.  
The relevant propagator combinations, putting Lorentz indices
back in, are
\begin{align}
\sum  s_m k^i k^j D_{ba}(\vec k) D_{bc}(\vec p) D_{ac}(\vec q) \bigg|_{\vec k = \vec q =-\vec p}
&= k^i k^j\eta^{\mu\nu} \! \bigg[
D_{\mu i} (\vec k) D_{0j}(\vec p) D_{0\nu}(\vec q) +
D_{\mu 0} (\vec k) D_{i0}(\vec p) D_{j\nu}(\vec q)
\nn \\ &-
D_{\mu 0} (\vec k) D_{ij}(\vec p) D_{0\nu}(\vec q)-
D_{\mu i} (\vec k) D_{00}(\vec p) D_{j\nu}(\vec q)
\bigg]_{\vec k = \vec q =-\vec p}
\nn \\
&= \sum C_{IJK}(\vec k) D_I(\vec k) D_J(\vec k) D_K  (\vec k) ,
\label{Cijk2}
\end{align}
and
\begin{align}
\sum s_m k^i k^j D_{ab}(\vec k) D_{bc}(\vec p) D_{ac}(\vec q) \bigg|_{\vec k = \vec p =-\vec q}
&=
k^i k^j\eta^{\mu\nu} \! \bigg[
D_{\mu i} (\vec k) D_{0j}(\vec p) D_{0\nu}(\vec q) +
D_{\mu 0} (\vec k) D_{i0}(\vec p) D_{j\nu}(\vec q)
\nn \\ &-
D_{\mu 0} (\vec k) D_{ij}(\vec p) D_{0\nu}(\vec q)-
D_{\mu i} (\vec k) D_{00}(\vec p) D_{j\nu}(\vec q)
\bigg]_{\vec k = \vec p =-\vec q}
\nn \\
&= \sum D_{IJK}(\vec k) D_I(\vec k) D_J(\vec k) D_K  (\vec k) ,
\label{Dijk2}
\end{align}
with $I,J,K = A,\xi$, and $C_{IJK},D_{IJK} \sim k^2$. 
Using $D_{I,ab}(\vec k) = (2\bar
\omega_I)^{-1} \e^{-i \bar \omega_I(\tau_a-\tau_b)}$
we have
\begin{align}
A_{3}^{\rm mix} &=
\frac{4}{\xi^2}\partial_{ \phi} 
m^2_A(\tau_a) \int_0^{\tau_a} \dd \tau_b \int_0^{\tau_a} \dd \tau_c \H (\tau_b)\H(\tau_c) 
\int \dbar^3 k 
 \sum 
\frac{1}{(2 \bar \omega_I) (2 \bar \omega_J) (2 \bar \omega_K) }
\label{A3a}
\\
& \qquad
\times \bigg[
 C_{IJK}\cos\big((\bar \omega_K-\bar \omega_I) \tau_a + (\bar \omega_I+\bar \omega_J)\tau_b -
(\bar \omega_J+\bar \omega_K)\tau_c \big)
\nn\\
&\qquad\qquad
- 2\Theta_{bc} D_{IJK}
\cos\big((\bar \omega_I+\bar \omega_K) \tau_a - (\bar \omega_I-\bar \omega_J)\tau_b -
(\bar \omega_J+\bar \omega_K)\tau_c\big) \bigg] .
\nn
\end{align}

Now use integration by parts with respect to $\tau_b$ and $\tau_c$ to write
\begin{align}
&\int_0^{\tau_a} \dd \tau_b \H(\tau_b) \int_0^{\tau_a} \dd \tau_c \H(\tau_c) 
\cos\[ (\bar \omega_I+\bar \omega_J)\tau_b -
(\bar \omega_J+\bar \omega_K)\tau_c+
(\bar \omega_K-\bar \omega_I) \tau_a \]
\nn \\
&\qquad =
 \frac{\H(\tau_a)^2}{(\bar \omega_J +\bar \omega_K)(\bar
  \omega_I+\bar \omega_J)} + ...
\end{align}
where we used the initial conditions \eref{BC1}.  And similarly
\begin{align}
&\int_0^{\tau_a} \dd \tau_b \H(\tau_b) \int_0^{\tau_a} \dd \tau_c \H(\tau_c) 
\Theta_{bc} 
\cos\[(\bar \omega_I+\bar \omega_K) \tau_a - (\bar \omega_I-\bar \omega_J)\tau_b -
(\bar \omega_J+\bar \omega_K)\tau_c\] 
\nn \\ &\qquad=
-\int_0^{\tau_a} \dd \tau_b \frac{\H(\tau_b)^2}{(\bar \omega_J +\bar \omega_K)}
\sin\[(\bar \omega_I+\bar \omega_K) \tau_a - (\bar \omega_I+\bar \omega_K)\tau_b\] 
\nn \\ &\qquad=
-\frac{\H(\tau_a)^2}{(\bar \omega_J +\bar \omega_K) (\bar \omega_I +\bar \omega_K)} .
\end{align}
The end result is
\begin{align}
A_{3a}^{\rm mix} &= \frac4{\xi^2} \partial_{ \phi} m^2_A(\tau)
\H^2(\tau) \int \dbar^3 k \sum
\frac{1}{(2\bar \omega_I)(2\bar \omega_J)(2\bar \omega_K)}
\nn \\
&\qquad\qquad \times
 \[ \frac{C_{IJK}}{(\bar \omega_I + \bar\omega_J)(\bar \omega_J+\bar \omega_K)}
+ \frac{2D_{IJK}}{(\bar \omega_J +\bar \omega_K) (\bar \omega_I +\bar
  \omega_K)}\]
\nn \\
&=\partial_{ \phi} m^2_A(\tau) 
\H^2(\tau) \frac{(1-3\xi-3\xi^2+\xi^3) - (1+\xi)^3}{64\pi^2\xi^2} \ln (\Lambda/\bar{m})^2 + \text{finite}
\nn \\
&=\partial_{ \phi} m^2_A(\tau) 
\H^2(\tau) \frac{-6(1+\xi)}{64\pi^2\xi} \ln (\Lambda/\bar{m})^2 + \text{finite} .
\label{A3amix}
\end{align}
%

\subsection{Summary of graphs}
\label{s:graph_summary}

In the previous subsections we have computed all quadratically and
logarithmically divergent contributions to the one-loop equation
of motion.  Here we collect and summarize the results, putting the
hats back on the relevant variables to indicate that we are still
in the conformal frame.

The first order graphs are given by \eref{A1mink}.
The second order contributions are (\ref{A2mink},\ref{A2mass},\ref{A2mix}),
and are summarized in Fig.~\ref{F:tadpole_result_2}.
At third order there is only one piece, given by \eref{A3amix}.
We now collect these terms into the three groups
$\hat A^{\rm Mink}$, $\hat A^{\rm mass}$ and $\hat A^{\rm mix}$.

The first and second order combined $\hat A^{\rm Mink}
=\hat A_1^{\rm Mink} + \hat A_2^{\rm Mink}$ is
\be
\hat A^\text{Mink}  =
\frac1{16\pi^2}\sum_\alpha S_\alpha \partial_{\hat \phi} \hat m^2_{\alpha} 
\[ \hat \Lambda^2 - \frac12 {\hat m}_\alpha^2 \ln \hat \Lambda^2 \]
+
\frac{(3+\xi)}{64\pi^2} \hat\lambda_{hA\theta}
\hat m^2_{A\theta}\ln (\hat\Lambda/\hat{\bar{m}})^2.
\label{Amink} 
\ee
As expected, this is independent of how the two-point interaction is split
into a free and
interacting term, since the first and second order pieces combined in the sum
$\hat m_\alpha^2 = \hat{\bar m}_\alpha^2 + \delta \hat m_\alpha^2$.
For $A_0$ mass insertions we have the second order piece \eref{A2mass}
\be
\hat A^\text{mass} = -\partial_{\hat \phi} \hat m_A^2 \delta \hat m_0^2
    \frac{3+\xi^2}{128\pi^2} \ln (\hat \Lambda/\hat{\bar{m}})^2 .
\label{Amass}
\ee
For the mixed piece we have contributions from second order \eref{A2mix}
and third order \eref{A3amix}, giving a total
\be
\hat A^\text{mix} = \partial_{\hat \phi} \hat m_A^2
  \( \frac{3\H'(1-\xi)^2}{\xi} - \frac{6\H^2(1+\xi)}{\xi} \)
  \frac{1}{64\pi^2} \ln (\hat \Lambda/\hat{\bar{m}})^2 .
\label{Amix}
\ee
All factors in (\ref{Amink}, \ref{Amass}, \ref{Amix}) that are
time-dependent~--- being the $\hat m^2$'s, $\hat \lambda_{hA\theta}$
and $\H$~--- are understood to be evaluated at $\tau$.

\begin{figure}[t]
\begin{center}
\begin{tikzpicture}
[line width=1.2 pt, scale=1.4]
\node at (-0.8,0) {$\sum A^{(2)}_i = \Bigg[$};
\begin{scope}[shift={(0,0)}]
\draw (0.0,0)--(1,0);
\draw (1.5,0) circle (0.5cm);
\draw[fill=black] (1,0) circle (.05cm);
\draw[fill=black] (2,0) circle (.05cm);
\node at (1.5,.7) {$D_h^{\alpha+}$};
\node at (1.5,-.7) {$D_h^{\alpha+}$};
\node at (1.0,-1.2) {$\partial_\phi \hat m_h^2 \delta \hat m^2_h$};
\end{scope}
\node at (2.5,0) {$+$};
\begin{scope}[shift={(3,0)}]
\draw (0.0,0)--(1,0);
\draw (1.5,0) circle (0.5cm);
\draw[fill=black] (1,0) circle (.05cm);
\draw[fill=black] (2,0) circle (.05cm);
\node at (1.5,.7) {$D_\theta^{\alpha+}$};
\node at (1.5,-.7) {$D_\theta^{\alpha+}$};
\node at (1.0,-1.2) {$\partial_\phi \hat m_\theta^2 \delta \hat m^2_\theta$};
\end{scope}
\node at (5.5,0) {$+$};
\begin{scope}[shift={(6,0)}]
\draw (0.0,0)--(1,0);
\draw (1.5,0) circle (0.5cm);
\draw[fill=black] (1,0) circle (.05cm);
\draw[fill=black] (2,0) circle (.05cm);
\node at (1.5,.7) {$D_\eta^{\alpha+}$};
\node at (1.5,-.7) {$D_\eta^{\alpha+}$};
\node at (1.0,-1.2) {$-2\partial_\phi \hat m_\eta^2 \delta \hat m^2_\eta$};
\end{scope}
\node at (-0.5,-3) {$+$};
\begin{scope}[shift={(0,-3)}]
\draw (0.0,0)--(1,0);
\draw (1.5,0) circle (0.5cm);
\draw[fill=black] (1,0) circle (.05cm);
\draw[fill=black] (2,0) circle (.05cm);
\node at (1.5,.7) {$D_{\mu\nu}^{\alpha+}$};
\node at (1.5,-.7) {$D_{\rho\sigma}^{\alpha+}$};
\node at (1.0,-1.2) {$\partial_\phi \hat m_A^2 \delta \hat m^2_A (3+\xi^2)$};
\end{scope}
\node at (2.5,-3) {$+$};
\begin{scope}[shift={(3,-3)}]
\draw (0.0,0)--(1,0);
\draw (1.5,0) circle (0.5cm);
\draw[fill=black] (1,0) circle (.05cm);
\draw[fill=black] (2,0) circle (.05cm);
\node at (1.5,.7) {$D_{0\mu}^{\alpha+}$};
\node at (1.5,-.7) {$D_{0\nu}^{\alpha+}$};
\node at (1.0,-1.2) {$\partial_\phi \hat m_A^2 \delta \hat m_0^2 \frac{3+\xi^2}{4}$};
\end{scope}
\node at (5.5,-3) {$+$};
\begin{scope}[shift={(6,-3)}]
\draw (0.0,0)--(1,0);
\draw (1.5,0) circle (0.5cm);
\draw[fill=black] (1,0) circle (.05cm);
\draw[fill=black] (2,0) circle (.05cm);
\node at (1.5,.7) {$D_{0\mu}^{\alpha+}$};
\node at (1.5,-.7) {$\partial_{(x_b)_i} D_{i\nu}^{\alpha+}$};
\node at (1.0,-1.2) {$-\partial_\phi \hat m_A^2 \frac{3\H'(\xi-1)^2}{2\xi}$};
\end{scope}
\node at (-0.5,-6) {$+$};
\begin{scope}[shift={(0,-6)}]
\draw (0.0,0)--(1,0);
\draw (1.5,0) circle (0.5cm);
\draw[fill=black] (1,0) circle (.05cm);
\draw[fill=black] (2,0) circle (.05cm);
\node at (1.5,.7) {$D_{\theta}^{\alpha+}$};
\node at (1.5,-.7) {$D_{00}^{\alpha+}$};
\node at (1.0,-1.2) {$-\hat \lambda_{h A\theta} \delta \hat m_{A\theta}^2 \frac{3+\xi}{2}$};
\end{scope}
\node at (3.5,-6) {$\Bigg] \times \frac{-1}{32\pi^2}\log\Lambda^2$};
\end{tikzpicture}
\caption{
\label{F:tadpole_result_2}
The second order tadpole diagrams and their corresponding
mathematical expression (below each graph).
These Feynman diagrams are in (conformal) coordinate space, with
the left and right vertices at $x_a$ and $x_b$ respectively.  The
argument of each of the propagators is $(x_b-x_a)$, and all time-dependent
quantities ($\hat \lambda_{A\theta}$, $\hat m^2_\alpha$,
$\delta \hat m^2_\alpha$ and $\H$) are evaluated at $\tau$.
}
\end{center}
\end{figure}

\section{Effective action}
\label{s:effective}

The previous section found the one-loop equation of motion.  The
corresponding loop-corrected effective action is the one which, upon
applying the Euler-Lagrange equations for $\phi$, yields the
loop-corrected equation of motion found in Sec.~\ref{s:eom}.  The
effective action is defined this way up to an arbitrary
field-independent constant, and an overall minus sign which is fixed
to obtain the correct sign kinetic terms.  Our working assumption is
that the background is fixed, i.e.\ $a(\tau)$ not a function of the
background field $\phi(\tau)$.

The classical action is defined by
\be
\Gamma^\text{cl} = \int \dd^3 x \, \dd \tau \; \hat\mcL^\text{cl},
\ee
where $\hat A^\text{cl}$ is the equation of motion following from the
Lagrangian density $\hat\mcL^\text{cl}$:
\be
\hat A^\text{cl}
  = \(\frac{\delta\hat\mcL^\text{cl}}{\delta\hat\phi'}\)'
  - \frac{\delta\hat\mcL^\text{cl}}{\delta\hat\phi} .
\label{euler_lag}
\ee
From \eref{eom_clas} we have that $\hat A^\text{cl}=\hat\lambda^+$,
so the classical action is
\begin{align}
\Gamma^{\rm cl} &=
\int \dd^3 x\dd \tau  \[-\frac12 \hat \phi \(\partial_\tau^2 -\frac{a''}{a} \)\hat \phi
-\hat V\]
\nn \\
&= \int \dd^3 x \dd \tau \sqrt{-g_{\rm conf}}\[ -\frac1{2a^2} \phi \(\partial_\tau^2 
+2 \frac{a'}{a} \partial_\tau \) \phi
-  V\] \nn \\
&=\int \dd^3 x \dd t \sqrt{-g_{\rm phys}}\[ -\frac1{2} \phi \(\partial_t^2 
+3 H\partial_t \) \phi
-  V\] ,
\end{align}
where we used that $\H' +\H^2 = a''/a$, and $H = \dot a/a$.  In the
second line we went to unhatted quantities, and in the third we changed
to physical time.  The measure in conformal coordinates is
$\sqrt{-g_{\rm conf}} =a^4$, and in physical coordinates
$\sqrt{-g_{\rm phys}}=a^3$.  Taking the Euler-Lagrange equation
from the last line, using the definition \eref{euler_lag},
we get the familiar FLRW equation of motion:
$\ddot \phi + 3H\dot \phi +V_\phi =0$.

Applying the same procedure to the quantum corrected equation of motion
gives us the quantum corrected effective action.  The relevant terms
at the level of the equations of motion are summarized in
Sec.~\ref{s:graph_summary}, and the one-loop correction to the
effective action is defined as
\be 
\Gamma^{1-{\rm loop}} = \int \dd^3 x \dd \tau
(\hat\mcL^{\rm Mink} + \hat\mcL^{\rm mass} +\hat\mcL^{\rm mix} +{\rm finite}) .
\ee
In determining $\hat\mcL$, we must be careful to use the same sign
convention in the Euler-Lagrange equation as we did above in
\eref{euler_lag}, to ensure the corrections to the effective potential
have the correct relative sign.

All but one term in $\hat A^\text{Mink}$ are polynomial, the exception
being the $\hat\lambda_{hA\theta}$ term.  For this term, the $\hat\phi$
dependent factors are
\be
\hat\lambda_{hA\theta} \hat m^2_{A\theta}
= 4g^2 \( -\hat\phi'' + \H'\hat\phi + \H^2 \hat\phi \).
\ee
This expression comes from a Lagrangian
\be
-\frac12 \hat m^4_{A\theta}
= -2g^2 \( \hat\phi'^2 - 2\H\hat\phi\hat\phi' + \H^2\hat\phi^2 \).
\label{non_poly_term}
\ee
For the rest of the terms in $\hat A^\text{Mink}$, which are polynomial
in $\hat\phi$, the corresponding action is found simply by integrating
with respect to $\hat\phi$, and then negating (since the
$\delta\hat\mcL/\delta\hat\phi$ term in \eref{euler_lag} comes with a
minus sign).
All terms in $\hat A^\text{mass}$ and $\hat A^\text{mix}$ are also
polynomial in $\hat\phi$, so can be similarly integrated.
Thus, from (\ref{Amink},~\ref{Amass},~\ref{Amix}), and using
\eref{non_poly_term}, we obtain
\begin{align}
\hat\mcL^{\rm Mink}&= -\frac{1}{16\pi^2} \sum_\alpha S_\alpha \(
 \hat m_\alpha^2  \hat  \Lambda^2
-\frac14 {\hat m}_\alpha^4\ln (\hat \Lambda/\hat{\bar{m}})^2\)
 -
\frac{(3+\xi)}{128\pi^2} \hat m^4_{A\theta}
 \ln (\hat \Lambda/\hat{\bar{m}})^2 ,
\nn \\
\hat\mcL^{\rm mass}&= 
-\frac{1}{64\pi^2}\hat m^2_A\ln (\hat \Lambda/\hat{\bar{m}})^2 
\( \frac{(3+\xi^2)}{\xi} (2\H^2-\H')
\),
\nn \\
\hat\mcL^{\rm mix}&=
 -\frac{1}{64\pi^2}\hat m^2_A\ln (\hat \Lambda/\hat{\bar{m}})^2 
\( \frac{3(1-\xi)^2}{\xi}\H'
-\frac{6(1+\xi)}{\xi} \H^2 \),
\end{align}
with $\alpha = \{h,\theta,\eta,A,\xi\}$, and $S_\alpha = \{1,1,-2,3,1\}$.  

Now take out a factor $\sqrt{-g}$, and write the hatted variables in
terms of their unhatted counterparts to go back to the
physical frame.  Use that $\H^2 = a^2 H^2$ and $\H' = a^2(\dot H +
H^2)$, and transform to physical time.  All terms are proportional to
the fourth power of a mass, hence we factor out an $a^4$.  The
result is
\be 
\Gamma^{1-{\rm loop}} = \int \dd^3 x \dd t \sqrt{-g}\(
\mcL^{\rm Mink} + \mcL^{\rm mass} +\mcL^{\rm mix} +{\rm finite}\) ,
\ee
with
\begin{align}
\mcL^{\rm  Mink}&= -\frac{1}{16\pi^2} \sum_\alpha S_\alpha \(
 m_\alpha^2  \Lambda^2
-\frac14 m_\alpha^4\ln (\Lambda/\bar{m})^2\)
 -
\frac{(3+\xi)}{128\pi^2}  m^4_{A\theta}
 \ln (\Lambda/\bar{m})^2 , \label{lmink}
 \\
\mcL^{\rm  mass}&= 
-\frac{1}{64\pi^2} m^2_A\ln (\Lambda/\bar{m})^2 
\( \frac{(3+\xi^2)}{\xi} (H^2-\dot H)\), \label{lmass}
 \\
\mcL^{\rm  mix}&=
 -\frac{1}{64\pi^2} m^2_A\ln ( \Lambda/\bar{m})^2 
\( \frac{3(1-\xi)^2}{\xi}(\dot H + H^2)
-\frac{6(1+\xi)}{\xi} H^2 \). \label{lmix}
\end{align}
Whereas $\hat \Lambda$ was a conformal cutoff on conformal three-momentum (equivalent to comoving momentum), $\Lambda$ is now a physical cutoff on physical three-momentum.

When we now plug in the FLRW-corrected two-point interactions \eref{mass},
we find for the total one-loop effective action (up to field-independent terms)
\begin{align}
\Gamma^{1-{\rm loop}} &= \frac{-1}{16\pi^2}
\int \dd^3 x \dd t \sqrt{-g}\Bigg[
\(V_{hh} + V_{\theta \theta} + 3 m_A^2\)\Lambda^2
\nn \\
&\qquad
- \biggl(
\(V_{hh} -\dot{H}-2H^2\)^2+\(V_{\theta\theta}-\dot{H}-2H^2\)^2 +3 m_A^4
\nn \\
&\qquad
+2\xi V_{\theta\theta}m_A^2-(6+2\xi)g^2\dot{\phi}^2+6 m_A^2 \(\dot{H}+2H^2\)
\biggr)
\frac{\ln (\Lambda/\bar{m})^2}{4}  \Bigg].
\label{offshell}
\end{align}
Recall that $m_A^2=g^2\phi^2$.
This result is still gauge variant, which was to be expected. Gauge invariance is only achieved on-shell. For a time-independent situation  ($\phi(t)={\rm const.}$) the Nielsen identities \cite{nielsen}
\be
\frac{\partial V_{\rm eff}}{\partial \xi} + \frac{\partial \phi}{\partial \xi}\frac{\partial V_{\rm eff}}{\partial \phi}=0
\ee
show that the effective potential is only gauge invariant when the background field is in a minimum of the potential. Here (just like in \cite{MP}) we want to use the time-dependent version of this statement: the effective potential is only gauge invariant when the background field satisfies its equation of motion.
Going on-shell enables us to rewrite in \eref{offshell} the term proportional
to $\dot\phi^2$.  This term originated from the mixed Goldstone-gauge boson
``mass'', the last term in \eref{lmink}, and can be transformed to:
\be
\int \dd^4 x \sqrt{-g} \, m_{A\theta}^4
= \int \dd^4 x \, 4g^2 a^3 \dot \phi^2
= \int \dd^4 x \sqrt{-g} \, 4g^2 \phi V_\phi
= \int \dd^4 x \sqrt{-g} \, 4m_A^2 V_{\theta\theta}.
\label{masstheta}
\ee
In the third step we integrate by parts and go on-shell.
The last step uses Goldstone's theorem (it exploits the fact that the
potential is a function of $\Phi \Phi^\dag$ \cite{Goldstone1, Goldstone2, MP}).

On-shell the result \eref{offshell} takes the form
\begin{align}
\Gamma^{1-{\rm loop}} &= \frac{-1}{16\pi^2}
\int \dd^3 x \dd t \sqrt{-g}\Bigg[
(V_{hh} + V_{\theta \theta} + 3 m_A^2)\Lambda^2
\nn \\
&\qquad -
\biggl(
\(V_{hh} -\dot{H}-2H^2\)^2+\(V_{\theta\theta}-\dot{H}-2H^2\)^2 +3 m_A^4
\nn \\
&\qquad
-6m_A^2\(V_{\theta\theta}-\dot{H}-2H^2 \)
\biggr) \frac{\ln (\Lambda/\bar{m})^2}{4}  \Bigg],
\end{align}
which is gauge invariant, as it should be. Introducing the notation
\be
\tilde{V}_{\alpha\alpha}\equiv
V_{\alpha\alpha}-\dot H-2H^2,
\label{shift}
\ee
we rewrite the final result (up to field-independent terms)
\be
\Gamma^{\rm 1-loop} \!=\! \frac{-1}{16\pi^2}
\!\int\! \dd^3 x \dd t \sqrt{-g}\Bigg[
(\tilde{V}_{hh} + \tilde{V}_{\theta \theta} + 3 m_A^2)\Lambda^2
- \(
\tilde{V}^2_{hh}+\tilde{V}^2_{\theta\theta}
+3 m_A^4 -6 \tilde{V}_{\theta\theta}  m_A^2
\) \frac{\ln (\Lambda/\bar{m})^2}{4}  \Bigg].
\label{result}
\ee
%

\subsection{Fermions and additional scalars}

It is straightforward to add fermions and additional scalars to the
calculation. If these fields are coupled to the Higgs field, and thus
have a $\phi$-dependent mass term, they will contribute to the
effective equation of motion for the background Higgs field $\phi(t)$
and to the effective action. 

We assume the extra scalars are in a basis with canonical
kinetic terms and have diagonal masses, and do not mix with $h$.
Similarly, we assume the extra fermions have diagonal masses.
It is easy to relax these assumptions and generalize the results.

In terms of Feynman diagrams, there are extra tadpole graphs with
the additional scalars and fermions running in the loop. The
calculation for additional scalars is analogous to that of the Higgs
fluctuations $h$ already done, with a contribution at first and second
order. The result is
\be 
\Gamma^{1-{\rm loop}}_\text{(scalar)} =-\frac{1}{16\pi^2}
\int \dd^3 x \dd t \sqrt{-g}\Bigg[
V_{\chi\chi} \Lambda^2
- \( V_{\chi\chi} - \dot H - 2 H^2 \)^2
\frac{\ln (\Lambda/\bar{m})^2}{4}  \Bigg],
\label{extra_scalar}
\ee
where $\chi$ is the additional real scalar and $V(\chi,\phi)$ its
potential.

Just as for the bosons, the tadpole diagrams with a fermion loop can
be mapped to the calculation for Minkowski space, except that the ``mass''
terms now depend on the FLRW scale factor.  To discuss fermions in
curved space-time, one has to use the vielbein formalism to transform
to a local Lorentz frame, where Lorentz transformations and spin-$\tfrac12$
particles are well defined.  The vielbeins are defined via
\be
g_{\mu\nu} = \eps^a_\mu \eps^b_\nu \eta_{ab},
\ee
with $\eps^a_\mu = a\delta^a_\mu$ for a conformal FLRW metric
\eref{metric}.  The gamma matrices are $\{\bar \gamma^\mu, \bar
\gamma^\nu\} = 2 g^{\mu\nu}$,  with  $\gamma^a = \eps^a_\mu
\bar \gamma^\mu$ the usual Minkowski gamma-matrices.
With this notation the fermionic action is \cite{greene}
\be
\mcL_f = \int \dd^4 x \sqrt{-g} \bar \psi (\bar \gamma^\mu \nabla_\mu - m)
\psi,
\ee
with the covariant derivative $\nabla_\mu = \partial_\mu +
\Omega_\mu$, and $\Omega_\mu = (1/4) \omega_{ab\mu} \gamma^a
\gamma^b$, with $\Omega_0 = 0$ and $\Omega_i = (1/2) (a'/a) \gamma^i
\gamma^0$ for the conformal FLRW metric.\footnote{If the fermions are
  charged under gauge groups, there will be an additional gauge
  connection. These extra terms do not affect the effective action
  for $\phi$, and for simplicity we leave them out.}
%
%
We rescale the fermion field $\hat \psi = a^{3/2} \psi$ and mass $\hat
m_\psi = a m_\psi$.  The Dirac equation then becomes
\be
\( i \gamma^\mu \partial_\mu - \hat m_\psi\) \hat \psi =0,
\ee
which is of the usual Minkowski form.  Hence the end result is the
Minkowski \cite{heitmannfermion} result but with the replacement
$m_\psi \to \hat m_\psi = am_\psi$ \cite{heitmannfermion}:
\begin{align}
\Gamma^{1-{\rm loop}}_\text{(fermion)}
&= \frac{1}{16\pi^2} \sum_f
\int \dd^3 x \dd \tau \[ \hat m_\psi^2 \hat \Lambda^2 - \frac14\( 
\hat m_\psi^4 + \hat m''_\psi \hat m_\psi\) \ln (\hat \Lambda/\hat{\bar{m}})^2 \]
 \nn \\
&= \frac{1}{16\pi^2} \sum_f
\int \dd^3 x \dd t \sqrt{-g} \bigg[ m_\psi^2  \Lambda^2
 \nn \\
&\qquad- \frac14\( 
m_\psi^4  +  m_\psi^2 \left(\frac{\ddot m_\psi+ 3 H \dot m_\psi}{m_\psi}+\dot{H}+2H^2\right)\) \ln (\Lambda/\bar{m})^2 \bigg].
\label{gamma_fermion}
\end{align}
The sum is over all fermionic degrees of freedom, which are two
(helicity) states for a Weyl fermion and four states for a Dirac
fermion.  The first line is the Minkowski result with the replacement
$m_\psi \to \hat m_\psi$. In the second line we went to physical
coordinates by factoring out an overall $a^4$ factor, and rewriting
the $\hat m''_\psi$ in terms of derivatives with respect to physical
time $t$. The first contribution to the logarithmic term incorporates the
expansion of the universe. The second contribution to the logarithmic term is
because the $\phi$ field is rolling, and is also present in Minkowski
space-time.  

Again we can simplify this result by going on-shell. For a fermion mass $m_\psi = \lambda \phi$ that is linear
in the Higgs field~--- which is the case for Yukawa interactions and also
for gaugino masses in supersymmetric theories~--- this gives
\be 
\Gamma^{1-{\rm loop}}_\text{(fermion)} 
=\frac{1}{16\pi^2} \sum_f
\int \dd^3 x \dd t \sqrt{-g} \[ m_\psi^2  \Lambda^2 - \frac14\( 
m_\psi^4 - m_\psi^2 \tilde{V}_{\theta\theta} \) \ln (\Lambda/\bar{m})^2 \].
\ee
Here we have used, again,  the background field equations and Goldstone's theorem. $\tilde{V}_{\theta\theta}$ was defined in \eref{shift}.

\section{Initial conditions}
\label{s:bogol}

Our interactions are time-dependent, and thus we needed to define
the split between a time-independent mass and a time-dependent
two-point interaction \eref{BC0}
\be
m^2_{\alpha\beta}(t) = \bar m^2_{\alpha\beta} + \delta
m^2_{\alpha\beta}(t),
\qquad
\delta
m^2_{\alpha\beta}(0)=0.
\label{BC2}
\ee
We furthermore chose initial conditions for $\phi(t)$ and $a(t)$ such
that the off-diagonal and Lorentz violating two-point interactions
vanished completely at the initial time \eref{BC1}.
These choices ensured the simplicity of the propagators.  They
also ensured the vanishing of the $t=0$ boundary terms coming from
integration by parts when evaluating the loop diagrams in
Sec.~\ref{s:eom}.  If these boundary terms did not vanish, they
would yield extra contributions to the final result, contributions
that depend on the initial conditions, and that diverge as
$t \to 0$.

Our chosen initial conditions are peculiar, and are not the ones to
be used in a realistic situation.  The problem in straightforwardly
generalizing our calculation to arbitrary initial conditions are the
two-point interactions $(m^2)^{0i}, m^2_{00}, m^2_{\theta A}$.  To
simplify the structure of the free action, and use the standard
expressions for the propagator, we have treated them as interactions
$m^2_{\alpha\beta} = \delta m^2_{\alpha\beta}$.  To satisfy \eref{BC2}
then requires the initial conditions \eref{BC1}.

However, in principle there is nothing to stop us from also splitting these
two-point interactions into a free and interacting part, as in \eref{BC2}.
Technically, this is complicated, as Lorentz symmetry is broken, and
the gauge fields and Goldstone bosons all mix at the initial
time.  Nevertheless, in principle we can expand all fields in mode
functions, where the mode functions satisfy the off-diagonal mode
equations (diagonalizing the equations will result in a momentum-dependent
diagonalization).  Then \eref{BC2} is satisfied, all terms depending
on the initial conditions vanish, and the results are the same as for
our choice of initial conditions \eref{BC1}.

In slightly different words, we argue that the result is
independent of the initial conditions as long as we choose the initial
vacuum to be that of the free theory, which is defined by the split of the
quadratic term into a time-independent mass and a time-dependent
interaction term.  That is, solve the mode equations derived from
the free action with $\bar m_{\alpha \beta}^2$, and the corresponding
annihilation operators annihilate the vacuum.
The different vacua, corresponding to different initial conditions, are
then related by a Bogoliubov transformation.  For the scalar field theory
this was shown by Baacke et al.\cite{heitmanninitial} (see also
\cite{Baacke:1999ia}).  For U(1) model we require a more general
Bogoliubov transformation, with momentum and polarization dependent
coefficients that mix the fields.  In principle this should be
straightforward, but we will not present any further details here.

In practice, choosing the initial conditions \eref{BC1} simplifies
the calculation of the free field mode functions and propagators, and
eliminates boundary terms, which is why we choose it.  We have
argued that a full treatment of initial conditions would yield the
same result, at least for the divergent corrections to the equation
of motion.


\section{Conclusion}
\label{s:concl}

In this paper we have computed the one-loop divergent corrections to
the effective action of a U(1) charged scalar, whose background vacuum
expectation value is changing with time, and when the background
space-time is of FLRW form.  We used the in-in formalism and $R_\xi$
gauge, and our main aim was to demonstrate gauge invariance of the
one-loop corrections.  The gauge invariance is only manifest upon
using the equations of motion, i.e.\ on-shell, in accordance with the
Nielsen identities.  Our result is given by \eref{result}, and
directly generalizes a previous work \cite{MP}.  In comparison with
the result in a Minkowski background, one can obtain the FLRW
correction by shifting all scalar masses by $2H^2 + \dot H$.  We
showed that additional scalars running in the loop can be easily
accommodated \eref{extra_scalar}, a result which also mimics the
Minkowski case, but with a shifted mass.  For fermions in the loop,
the additional correction is \eref{gamma_fermion}.  Although our
assumptions for the initial conditions of the background scalar and
scale factor are unrealistic, we argued that, if handled in a more
general way, the result would be the same.

The computed correction can now be used to derive the renormalization
group and find the RG improved action, a task we defer to a future
publication.  An additional task left for the future is to take into
full account the backreaction of the scalar on space-time.
Essentially, one must allow for spin-0 fluctuations of the metric,
determine their mixing with the scalar, diagonalize to a new basis,
and use this basis as the starting point of the calculation.  A
further generalization is to include a non-minimal coupling to gravity,
so as to describe models of Higgs inflation.
Finally, one could also generalize the
decomposition of $\Phi$ \eref{phi_decompose} to allow for a
time-dependent classical background in the imaginary direction.

\section*{Acknowledgments}

This research was supported by the Netherlands Foundation for
Fundamental Research of Matter (FOM) and the Netherlands
Organisation for Scientific Research (NWO).
SM would like to thank Andrei Linde for illuminating discussions.

\appendix

\section{Action in detail}
\label{A:action}

Here we work out the explicit form of the action \eref{action}
to fourth order in quantum fluctuations.
Using conformal coordinates the overall volume factor is
$\sqrt{-g} = a^4$, and $g^{\mu\nu}=a^{-2}\eta^{\mu\nu}$.  Unless
otherwise stated, all indices below are raised and lowered using
the Minkowski metric.

Start with the kinetic term for the gauge field.  The connection
cancels in the field strength $F_{\mu\nu} =\nabla_\mu A_\nu -
\nabla_\nu A_\mu = \partial_\mu A_\nu -\partial_{\nu } A_\mu$, and
thus
\be
-\frac14 \int \dd^4 x \sqrt{-g}  F^2 = 
-\frac14 \int \dd^4 x F^{\mu \nu} F_{\mu \nu}= 
\frac12 \int \dd^4 x A_\mu \(\partial_\rho \partial^\rho \eta
^{\mu\nu} - \partial^\mu \partial^\nu\) A_\nu.
\label{F2}
\ee
As expected, the result is invariant under a conformal transformation
of the metric. The kinetic terms and potential for the Higgs field are
expanded as
\begin{align}
& \int \dd^4 x \sqrt{-g}  ( |D\Phi|^2 -V)
\nn \\
&= 
\int \dd^4 x \Bigg\{ \frac{a^2}{2} \bigg[
\sum_{\vp = \phi_R,\theta}  \(  (\partial \vp)^2
+g^2 A^2 \vp^2 \)  
+ 2 g A^\mu(-\theta \partial_\mu \phi_R 
+ \phi_R \partial_\mu \theta) \bigg] - a^4 V \Bigg\} \nn \\
&=
\int \dd^4 x \Bigg\{ \sum_{\vp = \phi_R, \theta} \[ -\frac12 \bigg(
\hat \vp ( \partial^2 - \frac{a''}{a} + a^2  V_{\vp  \vp} )\hat \vp
- g^2 A^2 \hat \vp^2 \bigg)
- \frac1{3!} a V_{\vp\vp\vp}
\hat \vp^3- \frac1{4!}  V_{\vp\vp\vp\vp} \hat \vp^4 \]
\nn \\
&\hspace{3cm}
+ g A^\mu\[-a \hat \theta \partial_\mu \(\frac{\hat \phi_R}{a}\) 
+ a \hat \phi_R \partial_\mu \(\frac{\hat\theta}{a}\) \]
\Bigg\} ,
\label{LDphi}
\end{align}
with $\phi_R = \phi+h$. The prime denotes derivative with respect
to conformal time $\tau$.  We rescaled the scalar fields
$\vp_\alpha = \{\phi,h,\theta\}$ as in \eref{hat}.

The gauge fixing action is
\be
S_{\rm GF}=
-\frac1{2\xi} 
\int \dd^4 x \sqrt{-g}\[ (g^{\mu\nu}\nabla_\mu A_\nu)^2 - (g^{\mu \nu} \nabla_\mu  A_\nu) 2\xi
g\phi_R \theta + \xi^2 g^2 \phi_R^2 \theta^2\].
\label{SGF}
\ee
The first term becomes
\begin{align}
\frac{-1}{2\xi} \int \dd^4 x (\partial_\mu A^\mu - \eta^{\mu\nu} \Gamma^\rho_{\mu \nu}
A_\rho)^2
&=\frac1{2\xi} \int \dd^4 x  A_\mu \[\partial^\mu \partial^\nu 
+2 \eta^{\alpha\beta} \Gamma^\mu_{\alpha \beta} \partial^\nu 
- \eta^{\alpha\beta} \Gamma^\mu_{\alpha\beta}
\eta^{\rho\sigma}\Gamma^\nu_{\rho\sigma}
\]A_\nu 
\nn \\
&=\frac1{\xi} \int \dd^4 x  \[
\frac12A_\mu \partial^\mu \partial^\nu A_\nu 
+A_0 \( \H' -2 \H^2\) A_0
-A_0 {2} \H \partial^iA_i
\].
\end{align}
To get the second line we used the explicit form of the connections
\eref{Gamma} to write
\be
A_\mu \eta^{\rho\sigma} \Gamma_{\rho\sigma}^\mu =
A_0 (\eta^{00} \Gamma^0_{00} +\eta^{ii}\Gamma^0_{ii})
= A_0 \H (1-3) = -2\H A_0 ,
\ee
and integration by parts
\be
4\int \dd^4 x A_0 \H \partial_0 A_0 = 
-2 \int \dd^4x A_0\H' A_0 .
\ee

The second term in \eref{SGF} we can partially integrate using
(with $\tilde A^\mu \equiv g^{\mu\nu} A_\nu$ to indicate the index is
raised with $g^{\mu\nu}$)
\be
\int \dd^4 x \sqrt{-g} (\nabla_\mu \tilde A^\mu) B = - \int \dd^4 x \sqrt{-g}
\tilde A^\mu \partial_\mu B .
\ee
This follows from the fact that we have a covariant volume and a
covariant derivative.  Note also that $\nabla_\mu g_{\mu \nu} =0$,
and it is therefore irrelevant whether the raised index is on $A$
or on $\nabla$.  Thus the second term in \eref{SGF} can be written as
\be
- \int \dd^4 x \, a^2 g A^\mu \( \theta \partial_\mu \phi_R +
\phi_R \partial_\mu \theta\)
= -\int \dd^4 x g A^\mu \[ 
a \hat \theta \partial_\mu \(\frac{\hat \phi_R}{a}\) 
+ a \hat \phi_R \partial_\mu \(\frac{\hat \theta}{a}\) \].
\ee 
The second term above will cancel with the last term in \eref{LDphi}.  The
complete gauge-fixing term is
\begin{align}
S_{\rm GF}  &= \int \dd^4 x\bigg\{ 
\frac1{\xi} \[
\frac12A_\mu \partial^\mu \partial^\nu A_\nu 
+A_0 \( \H' -2 \H^2\) A_0
-A_0 {2} \H \partial^iA_i\]
\nn \\ & -g A^\mu \[ 
\hat \theta \(\partial_\mu - \frac{a'}{a} \delta_\mu^0 \) \hat \phi_R 
+\hat \phi_R \(\partial_\mu - \frac{a'}{a} \delta_\mu^0 \) \hat \theta \]
-\frac12 \xi g^2 \hat \phi_R^2 \hat \theta^2
\bigg\}.
\end{align}

Finally the Faddeev-Popov term is
\be
S_{\rm FP} = \int \dd^4 x a^4 {\bar \eta} \[- \nabla^2 + \xi g^2 (\theta^2 -
\phi_R^2) \] \eta,
\label{SFP}
\ee
which follows from 
\be
\delta_\alpha G =-\frac{1}{g} \partial^2 \alpha -\frac{1}{g}
\Gamma^\mu_{\mu \rho} \partial^\rho \alpha + \xi g(\theta^2 -
\phi_R^2) \alpha = -\frac1{g} \nabla^2 \alpha + \xi g(\theta^2 -
\phi_R^2) \alpha ,
\ee
where we used $\delta_\alpha \phi_R = - \alpha \theta$, $\delta_\alpha
\theta = \alpha \phi_R$, and $\delta_\alpha A_\mu = (-1/g)\partial_\mu
\alpha$.  Use $\Gamma^\mu_{\mu \rho} = \partial_\rho
\sqrt{-g}/\sqrt{-g}$ to write the first term in \eref{SFP} as
\be
-\int \dd^4 x \sqrt{-g} {\bar \eta} \nabla^2 \eta  = 
-\int \dd^4 x \sqrt{-g} {\bar \eta} \( \frac{1}{\sqrt{-g}} \partial_\mu
\sqrt{-g} g^{\mu\nu} \partial_\nu \) \eta =
-\int \dd^4 x \hat {\bar \eta} \[\partial^2 - \frac{a''}{a}\] \hat \eta,
\ee
where in the last step we rescaled the anti-commuting scalars $\hat
{\bar \eta} = a {\bar \eta}$.  Hence
\be
S_{\rm FP} = -\int \dd^4 x \hat {\bar \eta} \[\partial^2 - \frac{a''}{a}
+\xi g^2(\hat \phi_R^2 -\hat \theta^2)
\] \hat \eta.
\ee

Putting it all together, we write the action as $S = \sum S^{(i)}$
with $i$ denoting the number of quantum fields each term in $S^{(i)}$
contains. Then
\begin{align}
&\hspace{-0.5em} S^{(0)} =
    \int\! \dd^4x \bigg\{\frac12 (\hat \phi')^2
    +\frac12\frac{a''}{a} \hat \phi^2 - a^4 V \bigg\} ,
    \label{S0} \\
&\hspace{-0.5em} S^{(1)} =
    \int\! \dd^4x \bigg\{ 
    -\hat h \(  (\partial^2 - \frac{a''}{a} )\hat \phi +  a^3 V_{\phi}\)
    \bigg\},
    \label{S1} \\
&\hspace{-0.5em} S^{(2)} =
    \frac12 \int\! \dd^4 x \Bigg\{A_\mu \[ (\partial_\rho \partial^\rho+g^2 \hat \phi^2) \eta
    ^{\mu\nu} - (1-\frac1{\xi})\partial^\mu \partial^\nu   \]A_\nu
 \nn \\
&  \hspace{8em}
    +\frac1{\xi}\[A_0 \( \H' -2 \H^2\) A_0
-A_0 {2} \H \partial^iA_i\]
    \nn \\
&  \hspace{8em}-
    \hat \theta ( \partial^2 - \frac{a''}{a} + a^2  V_{\theta\theta}+\xi
    g^2 \hat \phi^2 )\hat \theta
    -4g A^0 \hat \theta \(\partial_\tau - \frac{a'}{a} \) \hat \phi
    \nn \\
&  \hspace{8em}-
    \hat h ( \partial^2 - \frac{a''}{a} + a^2  V_{hh} )\hat h
    -2\hat {\bar \eta} \[\partial^2 - \frac{a'}{a} +
    \xi g^2 \hat \phi^2
    \] \hat \eta
    \Bigg\},
    \label{S2}\\
&\hspace{-0.5em} S^{(3)} =
    \int\! \dd^4 x \Bigg\{
    \!\!-\! S_{\alpha\beta\gamma} a V_{\alpha\beta\gamma}
        \hat \vp_\alpha \hat \vp_\beta \hat \vp_\gamma
    -2g A^\mu \hat \theta \(\partial_\mu - \frac{a'}{a} \delta_\mu^0 \) \hat h
    +g^2 (A^2  - \xi \hat \theta^2 -2\xi \hat {\bar \eta} \hat \eta) \hat \phi \hat h 
    \Bigg\},
    \label{S3}\\
&\hspace{-0.5em} S^{(4)} =
    \int\! \dd^4 x \Bigg\{
    \!\!-\! S_{\alpha\beta\gamma\delta} V_{\alpha\beta\gamma\delta}
        \hat \vp_\alpha \hat \vp_\beta \hat \vp_\gamma \hat \vp_\delta
    +\frac12 g^2 A^2 (\hat h^2+\hat \theta^2)  
    - g^2 \xi \hat {\bar \eta} (\hat \theta^2 - \hat h^2) \hat \eta 
    -\frac12 g^2\xi \hat \theta^2 \hat h^2
    \Bigg\},
    \label{S4}
\end{align}
with $\vp_\alpha = \{h,\theta\}$, and $S_{\alpha\beta\gamma(\delta)}$
symmetry factors.  For a quartic Higgs potential
\be
a^4 V =a^4 \[ \frac12 m^2 \phi^2 + \frac{\lambda}{4!} \phi^4
\]=\frac12 \hat m^2 \hat \phi^2 + \frac{\lambda}{4!} \hat \phi^4.
\ee


\section{Tadpole method}
\label{A:tadpole}

In this appendix we give the derivation of the one-loop
quantum corrected equation of motion, starting from first principles
to determine the symmetry factors of the diagrams.  We use the in-in
formalism (see Sec.~\ref{s:propagators} and
\cite{ctpschwinger,ctpkeldysh,ctpjordan,Bakshi1,Bakshi2,ctpcalzetta,ctpweinberg})
where all quantum fields are doubled, and labeled by $\pm$
superscripts.  Expanding the Higgs field around the classical
background as in \eref{expand}, the the quantum corrected equations of
motion follow from the vanishing of the tadpole \cite{ctpweinberg}:
\be
\langle h^+(\tau,\vec x) \rangle = 0,
\ee
where $<\cdot>$ denotes the vacuum expectation value.
The vanishing of the $h^-$ component gives the same result, and does
not have to be considered separately.  We can write this as a path
integral expression:
\begin{align}
0 &= \langle h^+(y) \rangle= \int\! D \psi_\alpha^+ \D \psi_\alpha^- h^+(y)
\e^{iS_{\rm int}[\psi_\alpha^+]-iS_{\rm int}[\psi_\beta^-] } \nn \\
&= \!\! \int\! \D \psi_\alpha^+ \D \psi_\beta^- h^+(y)
\[1 + i \int \!\! \dd^4 x(\mcL_{\rm int}^+ \!-\! \mcL_{\rm int}^-) - \frac1{2!}   \int \!\! \dd^4
x(\mcL_{\rm int}^+ \!-\! \mcL_{\rm int}^-) \int \!\! \dd^4 x'(\mcL_{\rm int}^+ \!-\! \mcL_{\rm int}^-) + ... \]
\nn \\
&= -i \int \dd^4x D^{++}_{h}(y-x) A(x) ,
\label{eom_expansion}
\end{align}
with $\psi_\alpha$ running over all fields.
The equations of motion are then
\be
A(x) = 0 .
\ee
The relevant interactions are given in Sec.~\ref{s:action}. The
one-loop equations of motion can be calculated order by order in
perturbation theory, that is, ordered by the number of
insertions coming from $\mcL_{\rm int}$.
The higher order expectation values can be evaluated by
taking all possible Wick contractions.

\paragraph{First order.} At zeroth order there is no contribution
because $\langle h^+ \rangle =0$ in the free theory.  At first order
there is a classical and quantum contribution.  Starting with the
classical tree-level contribution:
\be
0 = -i \int \dd^4 x \langle h^+(y) h^+(x) \rangle \lambda_h^+ (x) = -i \int \dd^4x
D^{++}_{h}(y-x) \lambda_h^+ (x)
\quad \Rightarrow \quad A_{\rm cl}= \lambda_h^+ (x) =0 .
\ee
The first order one-loop quantum contribution has only diagonal two-point
insertions; the off-diagonal two-point interactions only enter at higher order.  Then
\begin{align}
0 &= -i \int \dd^4 x \langle h^+(y) h^+(x) \psi_\alpha^+ 
(x)^2 \rangle \frac12 \partial_\phi m^2_\alpha= -i \int \dd^4x
D^{++}_{h}(y-x) D^{++}_{\alpha}(0) \frac12 \partial_\phi m^2_\alpha
\nn \\
&\Rightarrow \quad A_1 =
 \frac12 \partial_\phi m^2_\alpha D_\alpha^{++}(0) ,
\label{a1}
\end{align}
as there is only one possible Wick contraction.  Here the subscript on
$A_n$ denotes the $n$th order contribution.

\paragraph{Second order.} 
For convenience, in the rest of this section we will drop the $\pm$
superscript. It should be kept in mind that at all times one should
sum over all possibilities, and the incoming $h$ propagator is always
$D_h^{++}(y-x)$. This summation is done explicitly in the main text.

Consider the second order contribution, with one two-point insertion. The
three-point vertex connecting to the incoming $h$ field can either be in
the first or second factor of $\mcL^{\rm int}$, which cancels the $1/2!$
in front (symmetry under $x \leftrightarrow x'$). Consider first two
diagonal two-point insertions, and the three-point vertex diagonal as well:
\begin{align}
&-\frac14\int \dd^4 x \dd^4 x'  \partial_\phi m_\alpha^2(x)
m^2_\beta (x')  \langle
h(y) h(x) \psi_\alpha (x)^2 \psi_\beta (x')^2 \rangle
\nn \\
&=
-i \int \dd^4 x D_h(y-x) \[-\frac{i}2  \partial_\phi m^2_{\alpha}(x)\int\dd^4 x' 
m_{\beta}^2(x') D_{\alpha\beta}(x-x') D_{\alpha\beta}(x'-x)\],
\label{a2diagonal}
\end{align}
since there are two possible Wick contractions. The part between the
square brackets is $A_2^{\rm diag}$.
For the Minkowski contribution with $m^2_{\theta A}$ we get
\begin{align}
&-\int \dd^4 x \dd^4 x'  2g
m^2_{\theta 0} (x')  \langle
h(y) \big[(\partial_\tau -\H(\tau))h(x)\big] A_0(x) \theta(x) A_0(x') \theta(x') \rangle
\label{a2theta} \\
&=
-i \int \dd^4 x \big[(\partial_\tau -\H(\tau))D_h(y-x)\big] \[-i  2g\int\dd^4 x' 
m_{\theta 0}^2(x') D_{00}(x-x') D_{\theta \theta}(x'-x)\]
\nn \\
&=
-i \int \dd^4 x D_h(y-x) \[-i  2g(-\partial_\tau -\H(\tau))\int\dd^4 x' 
m_{\theta 0}^2(x') D_{00}(x-x') D_{\theta \theta}(x'-x)\],
\nn
\end{align}
as there is only one possible Wick contraction. To get the last
expression we integrated by parts.  The $\theta A_i\partial^i
h$-vertex does not contribute (even at higher order and finite terms),
as going through the same steps, we get a result of the schematic form
$\partial^i A(t) =0$.

Finally, for the diagram with one $(m^2)^{0i}$
insertion we get
\begin{align}
&-\frac12\int \dd^4 x \dd^4 x'  \partial_\phi m_{\mu\mu}^2(x)(
m^2)^{0i} (x')  \langle
A_\mu(x)^2 A_0(x') A_i(x') \rangle
\nn \\
&=
-i \int \dd^4 x D_h(y-x) \[-i  \partial_\phi m^2_{\mu\mu}(x)\int\dd^4 x' 
(m^2)^{0i}(x') D_{\mu 0}(x-x') D_{\mu i}(x'-x)\],
\label{a2mix}
\end{align}
as there are two possible Wick contractions. Both Wick contractions
give the same result.

\paragraph{Third order.} 
At third order the diagram with two mixed mass $(m^2)^{0i}$
contributes to the divergent part in $A$.  There is a symmetry under
the interchange of the positions of the three vertices, $(x,x',x'')$,
and taking a definite order removes the factor $1/3!$ in
\eref{eom_expansion}. The result is
\begin{align}
& \!\!\!\! i \int \dd^4 x \dd^4 x' \dd^4 x''  \frac12
\partial_\phi m_{\mu\mu}^2(x)
(m^2)^{0i}(x') (m^2)^{0i}(x'') \langle h(y) h(x) A^\mu(x)^2 A^0(x')A^i(x')
A^0(x'') A^i(x'') \rangle
\nn\\
&=
- i\! \int \!\! \dd^4 x D_h(\! y\! -\! x\! ) 
\nn \\ & \hspace{1cm}
\times \[ -\frac{1}2 \partial_\phi m_{\mu\mu}^2(x) \int \dd^4 x' \dd^4 x''
m_{0i}^2(x') m_{0i}^2(x'') \sum D_{\mu \rho}(\! x\! -\! x'\! ) D_{\sigma
  \pi}(\! x'\! -\! x''\! ) D_{ \kappa\mu}(\! x''\! -\! x\! )\],
\label{a3a} 
\end{align}
where the sum is over the four possible Wick contractions \eref{indices}. 
The term between brackets is $A^{\rm mass}_3$.



\begin{thebibliography}{99}

\bibitem{birrell}
  N.~D.~Birrell and P.~C.~W.~Davies,
  Cambridge, Uk: Univ. Pr. ( 1982) 340p

\bibitem{candelas}
  P.~Candelas and D.~J.~Raine,
  Phys.\ Rev.\ D {\bf 12} (1975) 965.

\bibitem{ringwald1}
  A.~Ringwald,
  Annals Phys.\  {\bf 177} (1987) 129.

\bibitem{ringwald2} 
  A.~Ringwald,
  Z.\ Phys.\ C {\bf 34}, 481 (1987).

\bibitem{greene}
  P.~B.~Greene and L.~Kofman,
  Phys.\ Lett.\ B {\bf 448} (1999) 6
  [hep-ph/9807339].

\bibitem{heitmannfermion}
  J.~Baacke, K.~Heitmann, C.~Patzold,
  Phys.\ Rev.\  {\bf D58 } (1998)  125013.
  [hep-ph/9806205].

\bibitem{CW}
  S.~R.~Coleman, E.~J.~Weinberg,
  Phys.\ Rev.\  {\bf D7 } (1973)  1888-1910.

\bibitem{Bezrukov:2007ep}
  F.~L.~Bezrukov, M.~Shaposhnikov,
  Phys.\ Lett.\  {\bf B659 } (2008)  703-706.
  [arXiv:0710.3755 [hep-th]].

\bibitem{rachel2}
  R.~Jeannerot, S.~Khalil, G.~Lazarides and Q.~Shafi,
  JHEP {\bf 0010} (2000) 012
  [arXiv:hep-ph/0002151].
  
\bibitem{flatland}
  K.~Enqvist and A.~Mazumdar,
  Phys.\ Rept.\  {\bf 380} (2003) 99
  [hep-ph/0209244].

\bibitem{AD1}
  I.~Affleck and M.~Dine,
  Nucl.\ Phys.\ B {\bf 249} (1985) 361.
\bibitem{AD2}
  M.~Dine, L.~Randall and S.~D.~Thomas,
  Nucl.\ Phys.\ B {\bf 458} (1996) 291
  [hep-ph/9507453].
 
\bibitem{shore}
  G.~M.~Shore,
  Annals Phys.\  {\bf 128} (1980) 376.

\bibitem{allen}
  B.~Allen,
  Nucl.\ Phys.\ B {\bf 226} (1983) 228.

\bibitem{ishikawa}
  K.~Ishikawa,
  Phys.\ Rev.\ D {\bf 28} (1983) 2445.

\bibitem{garbrecht}
  B.~Garbrecht,
  Nucl.\ Phys.\ B {\bf 784}, 118 (2007)
  [hep-ph/0612011].

\bibitem{fukuda}
  R.~Fukuda and T.~Kugo,
  Phys.\ Rev.\ D {\bf 13} (1976) 3469.
 \bibitem{nielsen}
  N.~K.~Nielsen,
  Nucl.\ Phys.\ B {\bf 101} (1975) 173.

\bibitem{MP}
  S.~Mooij and M.~Postma,
  JCAP {\bf 1109}, 006 (2011)
  [arXiv:1104.4897 [hep-ph]].

\bibitem{heitmann1}
  J.~Baacke, K.~Heitmann and C.~Patzold,
  Phys.\ Rev.\  D {\bf 55} (1997) 7815
  [arXiv:hep-ph/9612264]. 

\bibitem{heitmann2}
  J.~Baacke, K.~Heitmann,
  Phys.\ Rev.\  {\bf D60 } (1999)  105037.
  [hep-th/9905201].

\bibitem{heitmann3}
  K.~Heitmann,
  Phys.\ Rev.\  {\bf D64 } (2001)  045003.
  [hep-ph/0101281].

\bibitem{boyanovsky}
  D.~Boyanovsky, D.~Brahm, R.~Holman and D.~S.~Lee,
  Phys.\ Rev.\  D {\bf 54} (1996) 1763
  [arXiv:hep-ph/9603337].

\bibitem{ctpschwinger}
  J.~S.~Schwinger,
  J.\ Math.\ Phys.\  {\bf 2 } (1961)  407-432.
  
\bibitem{ctpkeldysh}
  L.~V.~Keldysh,
  Zh.\ Eksp.\ Teor.\ Fiz.\  {\bf 47} (1964) 1515
  [Sov.\ Phys.\ JETP {\bf 20} (1965) 1018].

\bibitem{ctpjordan}
  R.~D.~Jordan,
  Phys.\ Rev.\  {\bf D33 } (1986)  444-454.
  
  \bibitem{Bakshi1}
  P.~M.~Bakshi, K.~T.~Mahanthappa,
  J.\ Math.\ Phys.\  {\bf 4 } (1963)  1-11.

\bibitem{Bakshi2}
  P.~M.~Bakshi, K.~T.~Mahanthappa,
  J.\ Math.\ Phys.\  {\bf 4 } (1963)  12-16.
  
\bibitem{ctpcalzetta}
  E.~Calzetta and B.~L.~Hu,
  Phys.\ Rev.\  D {\bf 35} (1987) 495
  
\bibitem{ctpweinberg}
  S.~Weinberg,
  Phys.\ Rev.\  {\bf D72 } (2005)  043514.
  [hep-th/0506236].
  
\bibitem{tadpole}
  S.~Weinberg,
  Phys.\ Rev.\ D {\bf 9} (1974) 3357.

\bibitem{heitmannmt}
K.~Heitmann,
Master's Thesis, 1996.

\bibitem{Heitmannphdt}
  K.~Heitmann,
  PhD Thesis, 2000.

\bibitem{heitmanninitial}
  J.~Baacke, K.~Heitmann, C.~Patzold,
  Phys.\ Rev.\  {\bf D57 } (1998)  6398-6405.
  [hep-th/9711144].

\bibitem{peskinschroeder}
  M.~E.~Peskin and D.~V.~Schroeder,
  Reading, USA: Addison-Wesley (1995) 842 p
  
 
\bibitem{GK}
  C.~Grosse-Knetter and R.~Kogerler,
  Phys.\ Rev.\ D {\bf 48} (1993) 2865
  [hep-ph/9212268].

  
\bibitem{Goldstone1}
  J.~Goldstone,
  Nuovo Cim.\  {\bf 19 } (1961)  154-164.
  
\bibitem{Goldstone2}
  J.~Goldstone, A.~Salam, S.~Weinberg,
  Phys.\ Rev.\  {\bf 127 } (1962)  965-970.
  
\bibitem{Baacke:1999ia} 
  J.~Baacke, D.~Boyanovsky and H.~J.~de Vega,
  Phys.\ Rev.\ D {\bf 63}, 045023 (2001)
  [hep-ph/9907337].


\end{thebibliography}
\end{document}